\documentclass{article}
\usepackage[utf8]{inputenc}
\usepackage{bbold}
\usepackage{hyperref}
\usepackage{xcolor} 
\usepackage{amsmath}
\usepackage{amssymb}
\usepackage{graphicx}
\usepackage{subcaption}
\usepackage{geometry}
\usepackage[thinlines]{easytable}
\usepackage[normalem]{ulem} 
\usepackage{feynmf}
\title{Breaking BEC}
\author{A.\;Kovtun$^{a,b}$\thanks{{\bf e-mail}:
akovtun@mpp.mpg.de}, M.\;Zantedeschi$^{a,b}$\thanks{{\bf e-mail}:
michaelz@mpp.mpg.de}
\\
$^a${\small{\em
Max-Planck-Institut f{\"u}r Physik,}}\\
{\small{\em F{\"o}hringer Ring 6, D-80805, Munich, Germany
}}\\
$^b${\small{\em Arnold Sommerfeld Center, Ludwig-Maximilians-Universit{\"a}t,
}}\\
{\small{\em Theresienstra{\ss}e 37, 80333, M{\"u}nchen, Germany
}}\\
} 
\date{March 2020}

\begin{document}

\maketitle
\begin{abstract}
   In this work quantum corrections to the classical evolution of a relativistic scalar condensate are studied. The problem is approached by means of two different perturbative approaches: the 2-particle-irreducible (2PI) effective action and the expansion in the self-coupling. In the weak coupling regime, the decoherence of the classical state is observed. The corresponding timescale is identified with the quantum break-time. 
\end{abstract}

\section{Introduction}
Among all possible configurations being studied in Quantum Field Theory (QFT), there is a class of solutions establishing the bond between quantum and classical field theory, namely the semi-classical non-perturbative solutions. The current understanding of these objects in QFT is somewhat problematic as there is no one-to-one correspondence between classical solutions and some states lying in the Hilbert space of the corresponding quantum theory. This naturally poses a question of the quantum fate of classical backgrounds when trying to understand their counterpart in QFT. Even if these solutions are classically stable, and hence long lived, we do not actually know how their quantum analogues evolve.

One way to address the problem of the quantum fate of semi-classical solutions has been addressed in \cite{Dvali:2012wq,Dvali:2012en,Dvali:2017eba} where explicit corpuscular models were built. In these works, the classical solution is explicitly constructed as a coherent state of weakly coupled free quanta, and, because this state is not an eigenstate of the full non-linear interacting hamiltonian, due to unitary evolution, the system will depart from its initial configuration. Since in these realisations the quanta are weakly coupled, perturbative estimates can be made regarding the time scale after which the system will deviate significantly from its initial coherent state structure. Such a timescale has been named quantum break time $t_{qb}$ \cite{Dvali:2012rt,Dvali:2017ruz} and it is generically set by the strength of the interaction coupling $\lambda$ and some classical timescale naturally embedded in the system
\begin{equation}
    \label{eq:quantumbreaktime}
    t_{qb}\sim \frac{t_{cl}}{\lambda}.
\end{equation}
Even though the coherent state point of view has proved fruitful as it gave interesting insights on many different topics such as black hole evaporation and information paradox, eternal inflation and cosmic axion \cite{Dvali:2012en,Dvali:2012rt,Dvali:2017ruz}, it only provides estimates for the quantum break time and, therefore, further investigation of this issue is needed. Hence, we are going to address the problem of the quantum fate of the classical backgrounds from a more familiar and scholastic point of view. We push the semiclassical treatment beyond the saddle point approximation by using the two-particle-irreducble (2PI) effective action.\footnote{The advantage of the semi-classical treatment, which is especially evident in light of the corpuscular approach, is that it does not rely on the corpuscular structure, though it still provides an instrument to capture some interesting quantum effects.} This effective action treats both the expectation value of the field and the connected propagator as independent variables and correspondingly takes into account their mutual interaction, thus allowing us to account for back-reactions on the background and vice-versa. Using this approach we can resolve the quantum dynamics of the system integrating the equations of motion derived as the stationarity condition of the 2PI effective action functional.  In order to obtain a real time evolution and observe the departure from the initial classical state (quantum breaking) explicitly, we integrate the equations of motion in the so called Schwinger-Keldysh (or \textit{in-in}) formalism. This choice  simplifies our task for two reasons: first of all, only the initial state needs to be specified as opposed to the $S-matrix$ formalism, where also an asymptotic out state is required. Secondly, solving the integro-differential set of equations become numerically affordable, as the evolution is guaranteed to be causal.

After describing the tools for studying the evolution of classical solutions, we are ready to specify a particular system we are going to work with. We consider a Bose condensate which is relevant in many branches of high energy and condensed matter physics (such as for corpuscular black holes \cite{Dvali:2017eba}, de Sitter \cite{Dvali:2012en}, dark matter description \cite{Sin:1992bg}, etc.). The model we are going to focus on is the simplest one: namely a homogeneous relativistic BEC in $1+1$ dimensions described by an $SO(2)$ symmetric scalar theory with a repulsive quartic self-interaction. In fact, for this system, it was possible to resolve explicitly its causal evolution as well as its departure from the classical trajectory.

Before describing the structure of this work let us briefly summarise our findings. Firstly, the loop expansion displays interesting features as the charge $Q$ of the condensate is varied. The quantum break time, in fact, seems to become asymptotically infinite when  $Q \rightarrow 0$, underlying the fact that the theory is closer and closer to its uncharged vacuum. Moreover, as we increase $Q$, for fixed coupling, the quantum break time, after reaching a minimum, starts growing again. We believe this is due to the high classicality of the mean-field solution in this limit. Secondly, we compared the 2PI effective action expansion with the perturbative expansion in the coupling. Although the two expansions seem to give the same real time evolution on small timescales, they drastically differ at later times. This fact comes as no surprise since the loop expansion of 2PI effective action resums much more diagrams (in fact, infinitely more) than the coupling expansion.

The article is organized as follows: in Section 2 we introduce the theory we are going to consider and the formalism for treating it. After that, in Section 3, we specify initial conditions for our evolutionary problem; in Section 4, we define the quantum breaking criterion and present the results of the numerical simulations. In Section 5 we present our conclusions.

\section{Theory and formalism}
In the following we set $\hbar=c=1$. However, we will indicate the working loop order in powers of $\hbar$.
\subsection{Classical theory}
Consider a scalar field theory with a repulsive quartic self-interaction in a $1+1$ dimensional finite box of size $L$ endowed with an $SO(2)$ global symmetry. The action of this theory is
\begin{equation}\label{eq:action}
    S[\varphi] = \int dx\, \left(\frac{1}{2}\left(\partial_\mu \varphi_a\right)^2 - \frac{1}{2} m^2 \varphi_{a}^{2}-\frac{\lambda}{16}\left(\varphi_{a}^{2}\right)^{2}\right),
\end{equation}
and is invariant under
\begin{equation}
    \varphi_a \rightarrow R_{ab}(\alpha)\varphi_b ,
\end{equation}
where 
\begin{equation}\label{eq:2d_rotation_matrix}
    R_{ab}(\alpha) = \left(
    \begin{array}{cc}
        \cos (\alpha) & \sin (\alpha)  \\
        -\sin (\alpha) & \cos (\alpha)
    \end{array}
    \right).
\end{equation}

Classically, the condensate solution is given by
\begin{equation}\label{eq:classical_ansatz}
 \varphi_{a}(t) = v\,R_{ab}(\omega t) \Vec{e}_b,
\end{equation}
with the modulus $v$ being
\begin{equation}\label{eq:modulus}
    v =\sqrt{\frac{4 \left(\omega^2-m^2\right)}{\lambda}}.
\end{equation}
Here, $\Vec{e}$ is a (constant) unit vector and $\omega$ is the frequency of isospin rotation covering the range $\omega \in (m,+\infty)$. 

The corresponding classical integrals of motion are
\begin{eqnarray}
    E =  \int dx\, \left(\frac{1}{2}\left(\partial_{t}\varphi_{a}\right)^2 + \frac{1}{2}\left(\partial_{x}\varphi_{a}\right)^2+ \frac{1}{2} m^2 \varphi_{a}^{2} + \frac{\lambda}{16}\left(\varphi_{a}^{2}\right)^{2}\right) = \nonumber\\
    \frac{L}{\lambda}\left(m^4-4 m^2 \omega^2 + \omega^4\right), \label{eq:classcial_energy} \\
    Q =  \int dx \left(\partial_{t}\varphi_{1} \varphi_{2} - \partial_{t}\varphi_{2} \varphi_{1}\right) = \frac{8\, \omega L \left(\omega^2 - m^2 \right)}{\lambda}. \label{eq:saddle_point_charge} 
\end{eqnarray}

We notice that the classical solution is an oscillating, homogeneous field having constant energy and charge, such that $E(\omega)\geq m Q(\omega)$. We have to point out that, classically, this field configuration is absolutely stable because it has no exponentially growing excitations in its spectrum; at the quantum level the same expectation might not be met. The reason is that field in the formulation given above is usually considered as a coherent state and, therefore, it is not a pure eigenstate of the Hamiltonian. Thus, evolution will scramble this state. The departure from its coherent structure is what we call quantum breaking.

\subsection{2PI effective action: generalities}
We are interested in studying the non-trivial time evolution of a Bose-Einstein-Condensate (BEC). For this purpose, the tools employed in non-equilibrium field theory are very advantageous. More specifically, we are going to use the effective action for composite operators. This framework has already proved its efficiency having been applied to studying different problems of non-equilibrium and statistical QFT's \cite{Berges:2004hn,Berges:2001fi}.

The formalism of the 2PI effective action was firstly introduced in \cite{deDominicis:1964zz,deDominicis:1964hh} and an efficient computational procedure was developed in \cite{Cornwall:1974vz}. Here we recap the main points to establish our notation.

Let us define the generating functional of the theory as
\begin{equation}
    Z[J,K] = \int \mathcal{D}\varphi_{a} \exp \left( i\left(S\left[\varphi_a\right] + \int \varphi_a (x)\,J_a (x) + \frac{1}{2}\int\int \varphi_a (x) K_{a b} (x,y)\varphi_b (y)\right) \right).
\end{equation}
Here, $J_a (x)$ and $K_{ab} (x,y)$ are the 1 and 2-particle sources.
    
The corresponding generating functional for connected diagrams is
\begin{equation}
    W[J,K] = - i \ln Z[J,K].
\end{equation}
    
To derive the 2PI effective action, we perform a Legendre transform w.r.t. both sources
\begin{equation}
    \Gamma\left[\phi,G\right] = W[J,K] - \int \frac{\delta W[J,K]}{\delta J_a (x)} J_a (x) -\frac{1}{2}\int\int \frac{\delta W[J,K]}{\delta K_{a b} (x,y)} K_{a b} (x,y),
\end{equation}
where we introduced
\begin{equation}
    \left\lbrace\begin{array}{rl}
       \displaystyle \frac{\delta W[J,K]}{\delta J_a (x)} = & \phi_a (x),   \\
       \\
    \displaystyle    \frac{\delta W[J,K]}{\delta K_{a b} (x,y)} = &  \displaystyle \frac{1}{2}\left( \phi_a (x) \phi_b (y) +  G_{a b} (x,y)\right),
    \end{array}\right.
\end{equation}
with $\phi_a (x)$ the expectation value of the field, and $G_{a b} (x,y)$ the full connected propagator of the theory.

In the limit of vanishing sources, one obtains the stationary conditions for the effective action
\begin{equation}
    \left\lbrace\begin{array}{c}
       \displaystyle \frac{\delta \Gamma[\phi,G]}{\delta J_a (x)} = 0,   \\
       \\
    \displaystyle    \frac{\delta \Gamma[\phi,G]}{\delta G_{a b} (x,y)} = 0.
    \end{array}\right.
\end{equation}
Solving these stationary conditions, accompanied with appropriate renormalization conditions, one can eventually compute the effective action or other quantities related to its functional derivatives.

It is possible to explicitly compute the 2PI effective action as the $\hbar$-expansion using \cite{Cornwall:1974vz}
\begin{equation}\label{eq:effective_action}
    \Gamma[\phi,G] = S[\phi] + \frac{i}{2}\mathrm{tr}\ln G^{-1} + \frac{i}{2} \mathrm{tr} \left( G_0^{-1}G \right) + \Gamma_2 [\phi,G],
\end{equation}
where $\Gamma_2 [\phi,G]$ is the sum of the vacuum-to-vacuum 2-particle irreducible diagrams computed with the following Feynman rules:
\begin{itemize}
   \item every internal line carries the propagator $G_{a b} (x,y)$
\item vertices are given by the non-linear part of the shifted action $S[\varphi_a+\phi_a]$
\end{itemize}
and
\begin{equation}\label{eq:G_0}
    G_{0,\,ab}^{-1} (x,y) = - i \frac{\delta^2 S[\phi]}{\delta {\phi}_a (x) \delta \phi_b (y)}.
\end{equation}

Summarizing all the definitions and imposing stationary conditions at the 2PI functional, the dynamical equations of motion for the 1 and 2-point Green's function can be obtained as

\begin{equation}\label{eq:stationary_conditions}
    \left\lbrace
    \begin{array}{c}
       \displaystyle \frac{\delta S[\phi]}{\delta \phi_a (x)} + \frac{i}{2} \mathrm{tr}\left( \frac{\delta G_0^{-1}}{\delta \phi_a (x)} G \right)  + \frac{\delta \Gamma_2 [\phi,G]}{\delta \phi_a (x)} = 0,
       \\
       \\
       \displaystyle   G_{ab}^{-1} (x,y) - G_{0\,ab}^{-1} (x,y) + 2 i \frac{\delta \Gamma_2[\phi,G]}{\delta G_{ab} (x,y)} = 0.
    \end{array}
    \right. 
\end{equation}
Note that to solve these equations a proper time-integration-contour must be specified. We are going to do this in the next paragraph. 

\subsection{\textit{In-in} contour}
As mentioned in the introduction, in order to obtain dynamical equations of motion that are causal, we choose the Schwinger-Keldysh time contour. Since the dynamics depends only on the past, we can easily simulate the difficult integral differential equations (\ref{eq:stationary_conditions}) using a finite difference scheme. It is simpler to evaluate the diagrams using the approach desribed in \cite{Berges:2004yj}.
Namely, we decompose the connected Green's function in two parts, $F$ and $\rho$, defined as
\begin{eqnarray}\label{eq:GF_decomposition}
    G_{ab}(x,y) = F_{ab}(x,y) - \frac{i}{2}\mathrm{sgn}_{\mathcal{C}}\left( x^0-y^0 \right) \rho_{ab}(x,y),\label{eq:G_decomp} \\
    F_{ab}(x,y) = \frac{1}{2}\left\langle \left\lbrace \phi_{a}(x), \phi_{b}(y)\right\rbrace \right\rangle ,\label{eq:F_def} \\
     \rho_{ab}(x,y) = i\left\langle \left[ \phi_{a}(x), \phi_{b}(y)\right] \right\rangle ,\label{eq:rho_def}
\end{eqnarray}
where the $\mathrm{sgn}_{\mathcal{C}}(x^0-y^0)$ is taken along the in-in time contour and guaranties proper time ordering. Here, $F_{ab}(x,y)$ is known as statistical propagator, while $\rho_{ab}(x,y)$ is related to the spectrum of the theory.

Substituting this decomposition in the second equation of (\ref{eq:stationary_conditions}) two equations for $F_{ab}(x,y)$ and $\rho_{ab}(x,y)$ are obtained. Thus, we have a set of three second-order differential-integral equations and three unknown functions (omitting counting w.r.t. field indices) $\phi_a,\, F_{ab}$ and $\, \rho_{ab}$. As we have already mentioned, to solve this system we have to complement it with a set of appropriate initial conditions. This will be done in Section 3.

\subsection{Generalized Ward-Takahashi identities as the cross-check criterion and the expansion validity}
As long as the theory we are considering is endowed with an $SO(2)$ symmetry, we have an efficient cross-check criterion to control the reliability of its evolution. This criterion is given by the generalized Ward--Takahashi identities for the 2PI effective action \cite{vanHees:2002bv}.

If the theory has a continuous global symmetry group $\mathcal{O}$ such that a field transformation $O \in \mathcal{O}$ leaves the Lagrangian invariant
\begin{equation*}
    \begin{array}{c}
         \displaystyle \phi_a \rightarrow O_{ab}\phi_b ,\quad O_{ab} = \mathbb{1}_{ab} + i\epsilon^{\alpha} \tau_{ab}^{\alpha} + \mathcal{O}\left(\epsilon^2\right),  \\
         \\
           \mathcal{L}\left(\phi_a,\partial\phi_a\right) = \mathcal{L}\left(O_{ab}\phi_b,\partial O_{ab}\phi_b\right),
    \end{array}
\end{equation*}
where $\tau_{ab}^{\alpha}$ are generators of the corresponding Lie algebra, then there is a set of integrals of motion given by
\begin{equation}
    \int_{\mathcal{C}} dx \, \frac{\delta \Gamma [\phi,G]}{\delta \phi_{a}(x)}\tau_{ab}^{\alpha}\phi_b (x) + \int_{\mathcal{C}}\int_{\mathcal{C}} dx\, dy\, \frac{\delta \Gamma [\phi,G]}{\delta G_{ab}(x,y)}\left(\tau_{ac}^{\alpha}\delta_{bd} + \tau_{ad}^{\alpha}\delta_{bc}\right) G_{cd}(x,y) = 0.
\end{equation}

Integrating by parts these identities one can extract the conserved currents. Here we do restrict ourselves to the case of $SO(2)$ symmetry, so we have only one generator and, correspondingly, a single conserved current given by
\begin{equation}\label{eq:general_charge}
    Q = \int d\mathbf{x}\, \lim_{y\rightarrow x} \epsilon_{ab}\partial_{x^0}\left(\vphantom{\left(lol\right)^2}\phi_{a}(x)\phi_{b}(y) +  G_{ab}(x,y)\right),
\end{equation}
where $\epsilon_{ab}$ is 2-dimensional Levi-Civita tensor and $d\mathbf{x}$ denotes integration over spatial coordinates. In terms of decomposition (\ref{eq:GF_decomposition}), the charge is
\begin{equation}
    Q = \int d\mathbf{x}\, \lim_{y\rightarrow x} \epsilon_{ab}\partial_{x^0}\left(\vphantom{\left(lol\right)^2}\phi_{a}(x)\phi_{b}(y)+  F_{ab}(x,y)\right).
\end{equation}

This quantity is conserved along the evolution of the system regardless of the given approximation. This turned out to be a useful criterion to check the reliability and stability of our simulations.

Another issue which has to be mentioned is the validity of the expansion. As long as we start from a semi-classical solution we have to know whether the expansion we are using is valid or not. We claim that this is the case because the theory is well-defined in the semi-classical limit. In fact, by means of rescaling
\begin{equation}
 \label{eq:resc}
    \phi_a \rightarrow \frac{m}{\sqrt{\lambda}}\phi_a, \quad
    x_{\mu} \rightarrow m x_{\mu},
\end{equation}
action \eqref{eq:action} becomes
\begin{equation}
\label{eq:dimensionact}
    S[\phi] = \frac{m^2}{\lambda} \Tilde{S}[\Tilde{\phi}],
\end{equation}
where $\Tilde{S}[\Tilde{\phi}]$ is dimensionless and independent of any coupling. Thus, we have a factor of $m^2 /\lambda$ in front of the dimensionless action validating the semi-classical approximation for small value of the coupling constant.

\section{BEC at one loop. Initial conditions}
To set up the numerical simulations, we have to provide appropriate initial conditions. Since we are interested in non-thermal configuration, we are not going to consider thermal distributions as for example in \cite{Berges:2004hn}. Hence, it seems natural to extrapolate initial conditions for $G_{ab}$ and $\phi_a$ from the 1-loop approximation of the effective action. 
Since at this order the considered diagrams are local, the dynamics of $G_{ab}$ and $\phi_a$ is easily solved. Thus, it is possible, exploiting the symmetry of the system, to calculate explicitly the asymptotic configuration within the $\hbar$ order.   

Let us consider the 2PI effective action at 1-loop level

\begin{equation}
    \Gamma [\phi,G] = S[\phi] + \frac{i}{2}\mathrm{tr}\left(\ln G^{-1} + G^{-1}_0 G\right)
    .
\end{equation}
The stationary conditions in this case are
\begin{equation}\label{eq:1loop_stationary_conditions}
   \left\lbrace \begin{array}{l}
  \displaystyle G^{-1}_{0,ab}(x,y) = G^{-1}_{ab}(x,y), \\
   \\
    \displaystyle \frac{\delta S}{\delta\phi_{a} (x)} +\frac{i}{2}\mathrm{tr}\frac{\delta  G^{-1}_0}{\delta\phi_{a}} G = 0,
    \end{array}
    \right.
\end{equation}
where, from definition (\ref{eq:G_0}) and action (\ref{eq:action}) it follows that

\begin{equation}
    G_{0, ab}^{-1}(x,y) = i\left(\left(\partial^2 + m^2 + \frac{\lambda}{4}\phi_{a}^{2}\right)\delta_{ab} + \frac{\lambda}{2}\phi_{a}\phi_{b} \right)\delta^{(2)}\left( x- y\right).
\end{equation}
One can notice that after evaluating the first stationary condition in (\ref{eq:1loop_stationary_conditions}), we can plug the result back into the effective action and derive
\begin{equation*}
    \Gamma [\phi,G[\phi]] = S[\phi] + \frac{i}{2}\mathrm{tr}\ln G_{0}^{-1} + \mathcal{O}\left(\hbar^2\right)
    \equiv \Gamma[\phi]_{1PI} + \mathcal{O}\left(\hbar^2\right)
    ,
\end{equation*}
which is the 1-particle irreducible effective action at 1 loop order.

The condensate solution of equations (\ref{eq:1loop_stationary_conditions}) is given by 
\begin{equation}\label{eq:}
\left\lbrace\begin{array}{l}
     \displaystyle \phi_a (t,x) = R_{ab}(\omega t) f_{b},  \\
     \\
    \displaystyle G_{ab}(t,x;\tau, y) =   R_{ac}(\omega t)R_{bd}(\omega \tau) \Tilde{G}_{cd}(t-\tau,x-y) ,
\end{array}\right. ,
\end{equation}
where $f_a$ is a constant vector, $R_{ab}(x,y)$ was defined in (\ref{eq:2d_rotation_matrix}), and $G_{ab}$ is 
\begin{equation}\label{eq:fourier_of_1loop_GF}
\begin{array}{l}
    \displaystyle\Tilde{G}_{ab}(t-\tau,x-y) = \frac{1}{L}\int \frac{d\gamma}{2\pi}\sum_{n=-\infty}^{+\infty}e^{-i\gamma (t-\tau)+i p_n (x-y)}\left.\Tilde{G}_{ab}(\gamma,p_n)\right|_{ p_n=\frac{2\pi n}{L}} ,\\
    \\
    \displaystyle\Tilde{G}_{ab}(\gamma,p_n) = \frac{\displaystyle i\left(\left(\displaystyle\omega^2+\gamma^2 - p_{n}^{2} - m^2 -\frac{\lambda}{4}f_{d}^{2} \right)\delta_{ac} + 2 i \omega\gamma \epsilon_{ac} - \frac{\lambda}{2}f_{a}f_{c}\right)}{\displaystyle\left(\vphantom{\frac{\lambda}{4}f_{d}^{2}}\gamma^2-\gamma_{+}^{2}(p_{n}) + i 0\right)\left(\vphantom{\frac{\lambda}{4}f_{d}^{2}}\gamma^2-\gamma_{-}^{2}(p_{n}) + i 0\right)}.
    \end{array}
\end{equation}
The quantities $f_a$ and $\gamma_{\pm}(p_n)$ are explicitly derived in Appendix \ref{sec:1PI_solution}.

Now, having the equilibrium Green's function under decomposition \eqref{eq:GF_decomposition}, we extract from (\ref{eq:fourier_of_1loop_GF}) its  statistical component
\begin{eqnarray}
    F^{0}_{ab}(t,x;\tau,y) = \frac{1}{L}\sum_{n=-\infty}^{+\infty} R(\omega t)\left( \frac{1}{4\gamma_{+}}\left( e^{-i\gamma_{+} (t-\tau)}\Tilde{G}(\gamma_{-},p_{n}) + e^{i\gamma_{+} (t-\tau)}\Tilde{G}(-\gamma_{+},p_{n}) \right) - \right. \nonumber
    \\
    \left. \frac{1}{4\gamma_{-}}\left( e^{-i\gamma_{-} (t-\tau)}\Tilde{G}(\gamma_{-},p_{n}) - e^{i\gamma_{-} (t-\tau)}\Tilde{G}(-\gamma_{-},p_{n}) \right) \right)R^{T}(\omega \tau)\frac{e^{i\,p_{n}(x-y)}}{\left(\gamma_{+}^{2}-\gamma_{-}^{2}\right)},
    \label{eq:F_initial}
\end{eqnarray}
which will serve as initial condition.

Let us now summarize all the initial conditions. For the field we have
\begin{equation}\label{eq:initial_conditions_for_phi}
    \left\lbrace
    \begin{array}{rl}
        \displaystyle \phi_{a} (0,x)  = & R_{ab}(0)f_{b},   \\
        \\
         \displaystyle \dot{\phi}_{a} (0,x)  = & \displaystyle  \lim_{t\rightarrow 0}\dot{R}_{ab} (\omega t) f_{b},
    \end{array}
    \right.
\end{equation}
where $f_{a}=(f,0)^{T}$ and $f$ is given by eq. (\ref{eq:1loop_modulus_equation}).
For $F_{ab}(x,y)$ we use
\begin{equation}\label{eq:initial_conditions_for_F}
    \left\lbrace
     \begin{array}{c}
     \displaystyle    F_{ab}(0,x;0,y) = F^{0}_{ab}(0,x;0,y),  \\
    \displaystyle\lim_{t\rightarrow 0} \partial_{t}F_{ab}(t,x;0,y) = \lim_{t\rightarrow 0} \partial_{t}F^{0}_{ab}(t,x;0,y), \\
   \displaystyle \lim_{\substack{t\rightarrow 0 \\ \scriptstyle \tau \rightarrow 0}} \partial_{t}\partial_{\tau}F_{ab}(t,x;\tau,y) = \lim_{\substack{t\rightarrow 0 \\ \scriptstyle \tau \rightarrow 0}} \partial_{t}\partial_{\tau}F^{0}_{ab}(t,x;\tau,y),
    \end{array}
    \right.
\end{equation}
where $F^{0}_{ab}(t,x;\tau,y)$ is given by (\ref{eq:F_initial}). Finally, the initial conditions for $\rho$ follow from the canonical commutation relations
\begin{equation}\label{eq:initial_conditions_for_rho}
     \left\lbrace
     \begin{array}{l}
        \displaystyle  \left[\phi_{a} (t,x), \dot{\phi}_{b} (t,y)\right] = - i \, \delta_{ab}\delta \left(x - y\right)  \\
        \\
        \displaystyle     \left[\dot{\phi}_{a} (t,x), \dot{\phi}_{b} (t,y)\right] = 0 \\
        \\
        \displaystyle    \left[\vphantom{\dot{\phi}_{b} (t,y)}\phi_{a} (t,x), \phi_{b} (t,y)\right] = 0 
     \end{array}
     \right. \Rightarrow
     \left\lbrace
     \begin{array}{rl}
        \displaystyle    \lim_{y^0\rightarrow x^0} \partial_{x^0}\rho_{ab} (x,y) = & \delta_{ab}\delta (x-y) \\
        \\
        \displaystyle  \lim_{y^0\rightarrow x^0} \partial_{x^0}\partial_{y^0}\rho_{ab} (x,y) = & 0    \\
        \\
        \displaystyle    \vphantom{\lim_{y^0\rightarrow x^0}} \rho_{ab} (x^0,x;x^0,y) = & 0
    \end{array}
    \right.    
\end{equation}

\section{Quantum break-time}

We are now going to solve the causal equations of motion given by the stationary conditions for the 2PI effective action (\ref{eq:stationary_conditions}) along the \textit{in-in} time contour, equipped with the aforementioned initial conditions (\ref{eq:initial_conditions_for_phi}), (\ref{eq:initial_conditions_for_F}) and (\ref{eq:initial_conditions_for_rho}).
\subsection{Diagrammatics}
Let us specify the approximation to the effective action we are going to use in our numerical analysis.

We will retain all terms up to order $\hbar^2$ in the effective action (\ref{eq:effective_action}). These terms are given by all two-particle irreducible vacuum-to-vacuum diagrams with less than three loops. The $\mathcal{O}\left(\hbar^2\right)$ part of the effective action is given by

\begin{equation*}
    \Gamma[\phi,G] = S[\phi] + \frac{i}{2}\mathrm{tr}\ln G^{-1} + \frac{i}{2} \mathrm{tr} \left( G_0^{-1}G \right) + \Gamma_{2}^{(2)} [\phi,G] 
    ,
\end{equation*}

where 
\begin{equation}\label{eq:vacuum_diagramms}
\centering
\Gamma_{2}^{(2)}[\phi,G] ~ = ~~~
\parbox{30mm}{
\begin{fmffile}{vacvac1}
\begin{fmfgraph}(70,70)
\fmftop{t0,t1}
\fmfbottom{b0,b1}
\fmf{phantom}{t0,v1,t1}
\fmf{phantom}{b0,v1,b1}
\fmffreeze
\fmf{plain,tension=0.8,right=270}{v1,v1}
\fmf{plain,tension=0.8,right}{v1,v1}
\fmfdot{v1}
\end{fmfgraph}
\end{fmffile}
}
+
\parbox{30mm}{
\begin{fmffile}{vacvac2}
\begin{fmfgraph}(60,60)
\fmfleft{i}
\fmfright{o}
\fmf{phantom,tension=5}{i,v1}
\fmf{phantom,tension=5}{v2,o}
\fmf{plain,left,tension=0.4}{v1,v2,v1}
\fmf{plain}{v1,v2}
\fmfdot{v1,v2}
\end{fmfgraph}
\end{fmffile}
}.
\end{equation}
Here, the upper index in $\Gamma_{2}^{(2)}[\phi,G]$ implies that only $\hbar^2$ order contributions (or equivalently two loops terms) are retained.

As one can see from \eqref{eq:vacuum_diagramms}, there are two diagrams. The first one is independent of $\phi_a$ and it just corrects the local self-energy. Thus, the inclusion of this single diagram would not lead to any interesting deviation in the evolution. The second diagram is the field-dependent one coming from the $\lambda\phi_a\varphi_a\varphi_{b}^{2}$ interaction and this is the term leading to a non-trivial evolution already at the 2-loop level and to what we call quantum breaking.

After the evaluation of the diagrams we get the $\hbar^2$ part of the effective action
\begin{eqnarray}
    \Gamma_{2}^{(2)} [\phi,G] = -\frac{ \lambda}{16}\int_{\mathcal{C}} dx\left( 2 G_{ab}(x,x) G_{ab}(x,x) + \left(G_{aa}(x,x)\right)^2 \right) + \nonumber\\
    \frac{i  \lambda^2}{16}\int_{\mathcal{C}} dx \int_{\mathcal{C}} dy \left( \vphantom{\left(G_{aa}(x,x)\right)^2}G_{ab}\left(x,y\right)\phi_{a}(x)\phi_{b}(y) G_{cd}(x,y)G_{cd}(x,y) \,+\right. \\
    \left. 2 \, \phi_{a}(x) G_{ab}(x,y) G_{bc}(x,y)G_{cd}(x,y) \phi_{d}(y) \vphantom{\left(G_{aa}(x,x)\right)^2}\right),
\end{eqnarray}
where $\mathcal{C}$ denotes the time integration contour.

We plug this into (\ref{eq:stationary_conditions}) and derive the causual closed set of equations for the Green's functions and field expectation values
\begin{equation}\label{eq:main_EOM}
    \left\lbrace
        \begin{array}{l}
            \displaystyle - \frac{\delta S[\phi]}{\delta \phi_{a}(x)}  + \frac{ \lambda}{4}\left(F_{cc}(x,x)\delta_{ab} + 2 F_{ab}(x,x) \vphantom{\left(G_{aa}(x,x)\right)^2} \right)\phi_{b}(x) = 
             \int\limits_{0}^{x^0}dy^0\int\limits_{0}^{L}dy^1\,\Sigma^{\phi}_{ab} (x,y)\phi_{b}(y), \\
           \\
            \displaystyle \left( \partial^{2} \delta_{ab} + M_{ab}^{2}\left( x \right)\right) F_{bc}(x,y) = \int\limits_{0}^{L} d z^1\left(
             \int\limits_{0}^{y^0} dz^0 \Sigma^{F}_{ab}(x,z)\rho_{bc}(z,y)  - \int\limits_{0}^{x^0} dz^0 \Sigma^{\rho}_{ab}(x,z)F_{bc}(z,y) \right), \\ 
           \\
           \displaystyle  \left( \partial^{2} \delta_{ab} + M_{ab}^{2}\left( x \right)\right) \rho_{bc}(x,y) =
             - \int\limits_{y^0}^{x^0} dz^0 \int\limits_{0}^{L} d z^1\Sigma^{\rho}_{ab}(x,z)\rho_{bc}(z,y) ,
        \end{array}
    \right.
\end{equation}
where the mass matrix $M_{ab}(x)$ is the local part of the self-energy defined as
\begin{equation}
    M_{ab}^{2}(x) = \left(m^2 +\delta m^2\right) \delta_{ab} + \frac{\lambda}{4} \left( \vphantom{\left(G_{aa}(x,x)\right)^2} \phi_{c}^{2}(x) +  F_{cc}(x,x) \right)\delta_{ab} + \frac{\lambda}{2} \left( \vphantom{\left(G_{aa}(x,x)\right)^2} \phi_{a}(x)\phi_{b}(x) +  F_{ab}(x,x)\right) .
\end{equation}
The functions $\Sigma^{\phi}_{ab}(x,y)$, $\Sigma^{\rho}_{ab}(x,y)$ and $\Sigma^{F}_{ab}(x,y)$ are explicitly derived in Appendix \ref{sec:non_loc_self_energy}.


To solve these equations in momentum space, one must introduce a momentum cutoff. Our choice of this quantity was simply dictated by the independence of the dynamics from it. We therefore increased the cutoff up to the value where no differences in the dynamics were observed. 
\subsection{Definition of Quantum break-time}
In order to understand what is the indication of quantum breaking, we will introduce two quantities which will help us to identify the moment at which this phenomenon takes place. We will name these quantities as \textit{classical} $Q_{cl}(t)$ and \textit{quantum} $Q_{q}(t)$ charges and define them as
\begin{align}
         \displaystyle Q_{cl}(x^0) = & \int\limits_{0}^{L} dx^1 \lim_{y\rightarrow x} \epsilon_{ab}\partial_{x^0}\phi_{a}(x)\phi_{b}(y) = \int\limits_{0}^{L} dx^1\left(\dot{\phi}_{1}(x)\phi_{2}(x)-\dot{\phi}_{2}(x)\phi_{1}(x)\right), \label{eq:classical_charge}\\
         Q_{q}(x^0) = &\int\limits_{0}^{L} dx^1 \lim_{y\rightarrow x} \epsilon_{ab}\partial_{x^0}F_{ab}(x,y).\label{eq:quantum charge}
\end{align}
The sum of these two charges is the full charge of the system, given by (\ref{eq:general_charge}). As one can see from  (\ref{eq:F_initial}) and (\ref{eq:initial_conditions_for_phi}), it follows that initially $Q_{cl}(0) \gg Q_{q}(0)$. It is then natural to define the quantum break-time as the time when the quantum and classical charges become comparable to each other, i.e.
\begin{equation}\label{eq:qunatum_breaking_critetion}
   \boxed{ Q_{q}\left(t_{qb}\right) \simeq Q_{cl}\left(t_{qb}\right)}
\end{equation}


\subsection{Numerical simulation}
In the following we discuss our numerical results.

Eq.s (\ref{eq:main_EOM}) accompanied with initial conditions (\ref{eq:initial_conditions_for_phi}), (\ref{eq:initial_conditions_for_F}) and (\ref{eq:initial_conditions_for_rho}), have been solved numerically using a Newtonian finite difference scheme for derivatives and a trapezoidal rule for memory integrals. Both the size of the box as well as the spacing and the time step have been varied to ensure the reliability of the simulation by verifying the total charge conservation. Moreover, we explicitly checked that for what follows, effects due to the box-size are irrelevant as long as $L \gg m^{-1}$; from now on we therefore set $L= 10\,  m^{-1}$.
 
Since it is possible to rescale the action as in \eqref{eq:resc}, we set $m=1$  and work in units of mass.  Then, in order to explore the properties of the quantum breaking two parameters can been varied: namely the coupling $\lambda$, fixing the interaction strength, and the condensate frequency $\omega$, determining the initial field amplitude. Moreover, it should be noted that at $t=0$ we have $Q_{cl}\gg Q_{q}$ as the first one scales as $\lambda^{-1}$ and their difference is of order $\mathcal{O}(\lambda^0)$, so taking this into account we will not distinguish the initial classical charge from the full charge.

\begin{figure}
 \centering
    \begin{subfigure}[b]{0.52\textwidth}
        \includegraphics[width=\textwidth]{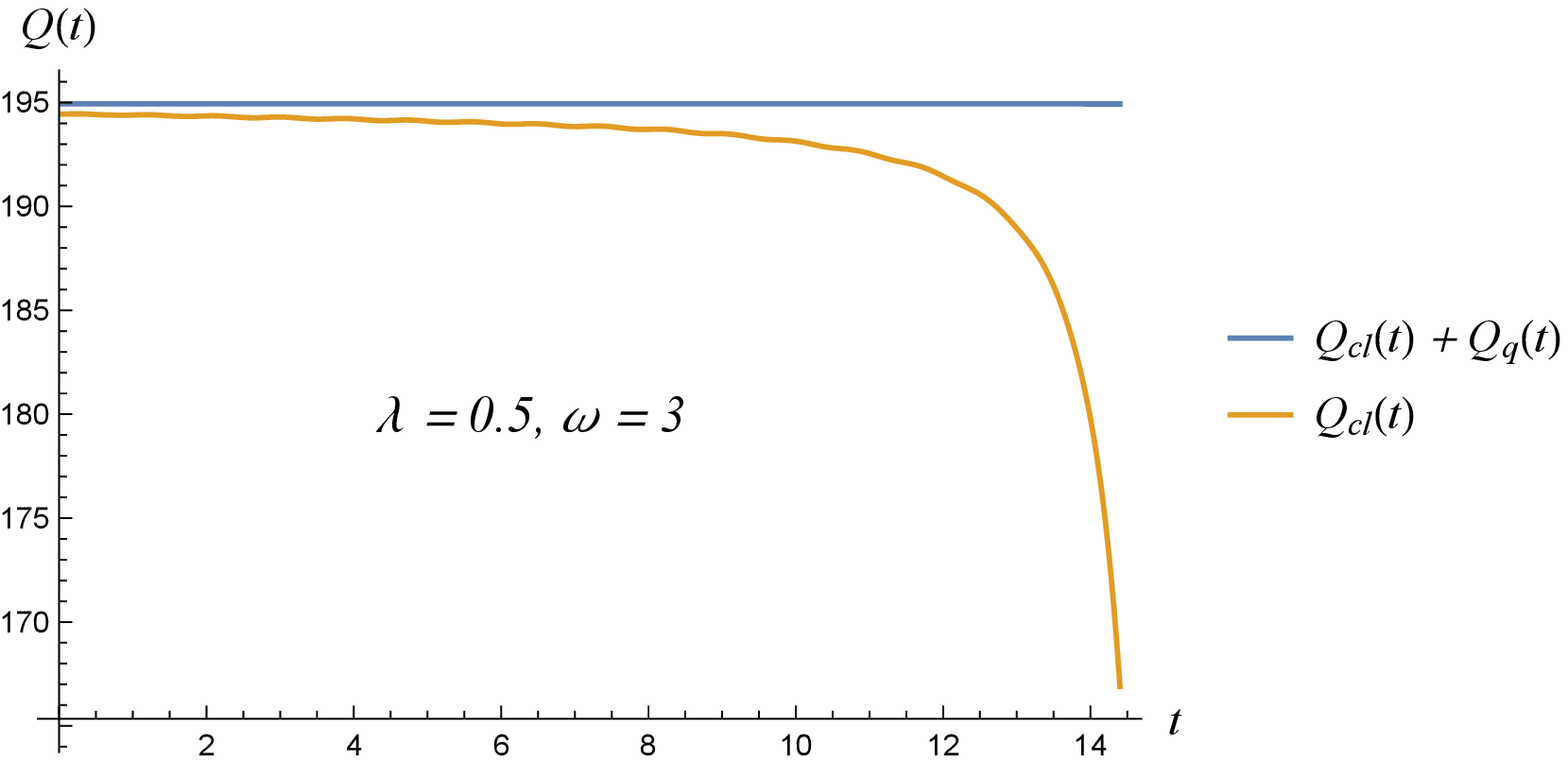}
        \caption{Time dependence of the classical and full charges for $\omega = 3$ and $\lambda = 0.5$.
    }
        \label{fig:Qcl_and_full_charges}
    \end{subfigure}
    ~
    \centering
    \begin{subfigure}[b]{0.47\textwidth}
        \includegraphics[width=\textwidth]{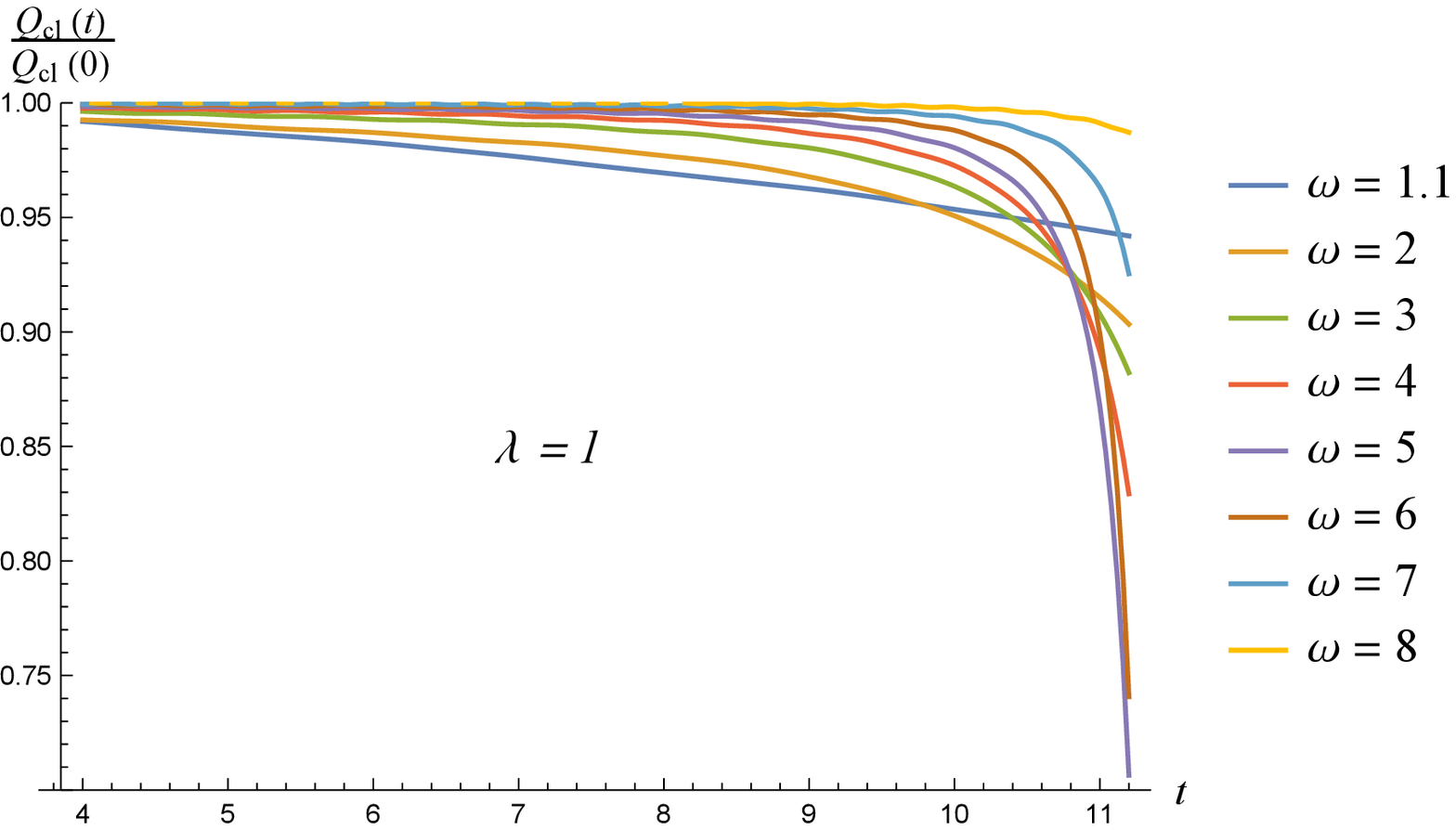}
        \caption{Time dependence of $Q_{cl}$ expressed in percents for different frequencies $\omega$.}
        \label{fig:Qcl(T)for_diff_omega}
    \end{subfigure}
    ~
    \centering
    \begin{subfigure}[b]{0.48\textwidth}
        \includegraphics[width=\textwidth]{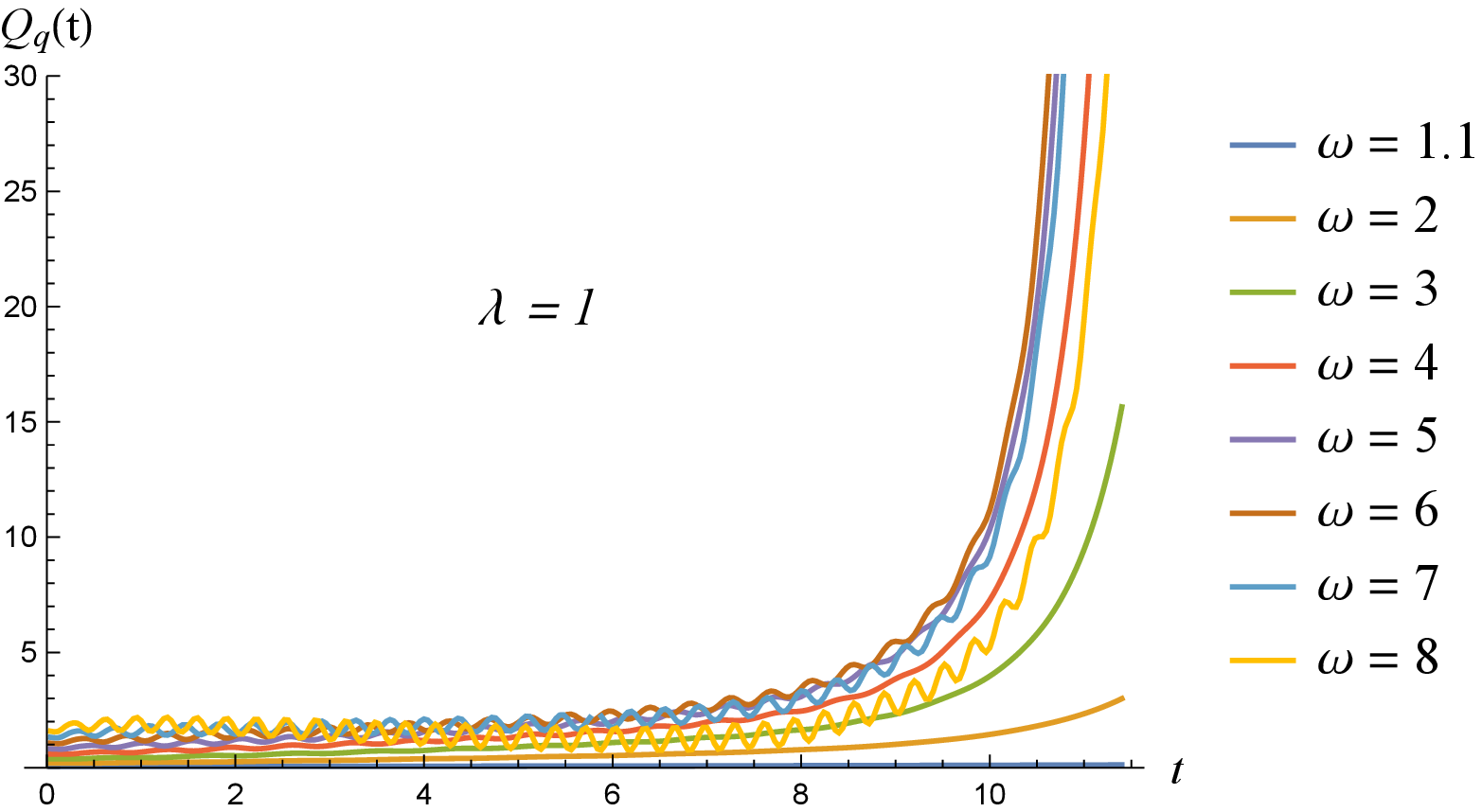}
        \caption{Time dependence of $Q_q$ for different frequencies $\omega$.}
        \label{fig:Qq(T)for_diff_omega}
    \end{subfigure}
    \caption{Time evolution of charges}
    \label{fig:Time_dependence_of_charge}
\end{figure}

The first results we want to present are showed in Fig. \ref{fig:Qcl_and_full_charges}. It is shown how the classical charge $Q_{cl}(t)$ decreases with time. This quantity is roughly constant during a rather long time scale (approximately 10 inverse masses), mimicking the classical tree level solution, and dramatically changes thereafter at an almost exponential speed;  this is the essence of quantum breaking. Moreover, the total charge (blue line) is shown to be conserved, therefore ensuring the reliability of the simulation. This feature holds true also in Fig. \ref{fig:Qcl(T)for_diff_omega}, although not displayed explicitly. Here, the dynamics of the breaking is shown for different $\omega$’s (or, equivalently, different charges) and, as one can see, they all seem to break in a similar fashion. For completeness, in fig. \ref{fig:Qq(T)for_diff_omega}, the evolution of the quantum charge defined according to (\ref{eq:quantum charge}) is shown for the same set of $\omega$’s. As expected, because of the charge conservation of the system, it grows with time. Since the occupation number of different modes can be extracted from the propagator $G$, we interpret the quantum breaking as a redistribution of such modes into a configuration where quantum fluctuations are no longer dynamically negligible.

\begin{figure}
    \centering
    \begin{subfigure}[b]{0.47\textwidth}
        \includegraphics[width=\textwidth]{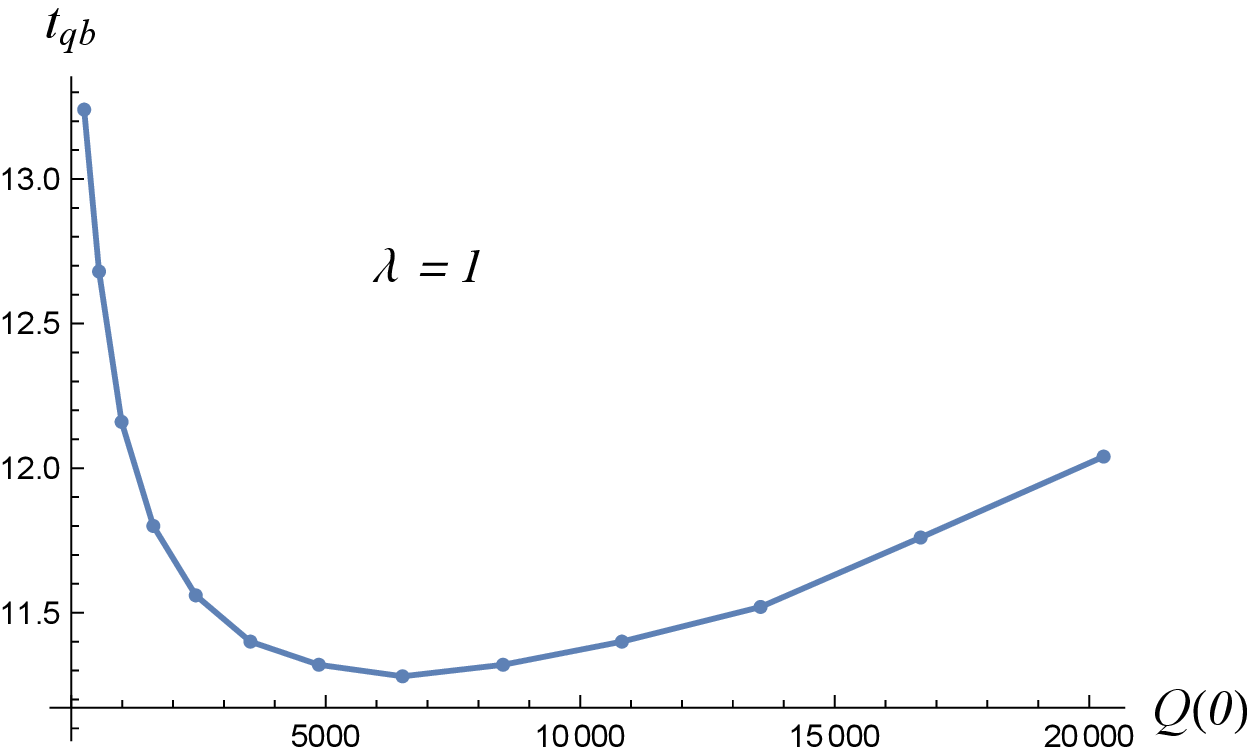}
        \caption{$\lambda = 1$}
        \label{fig:quantum_bt_vs_omega_lambda_1}
    \end{subfigure}
    ~
    \centering
    \begin{subfigure}[b]{0.47\textwidth}
        \includegraphics[width=\textwidth]{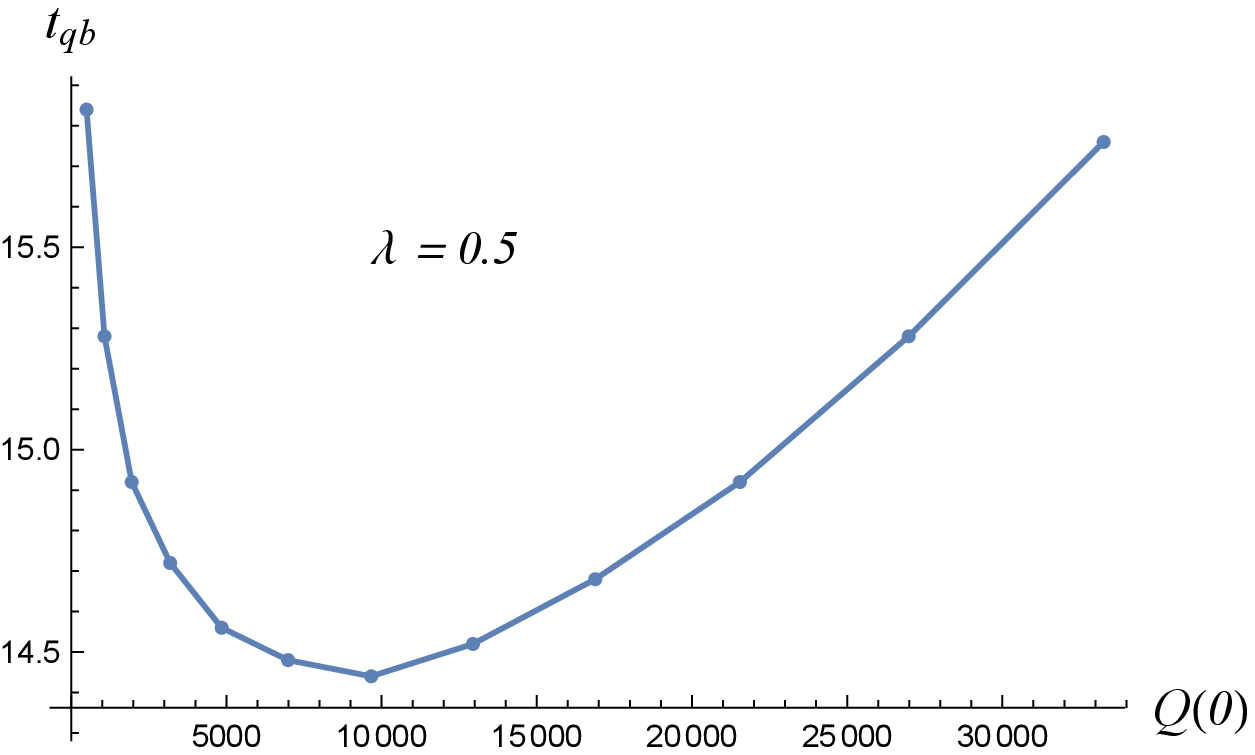}
        \caption{$\lambda = 0.5$}
        \label{fig:quantum_bt_vs_omega_lambda_05}
    \end{subfigure}
    \caption{Quantum break-time as a function of the full charge}
    \label{fig:quantum_break-time_vs_Q}
\end{figure}

The dependence of such “scrambling” on the frequency $\omega$ is explicitly depicted in Fig. \ref{fig:quantum_bt_vs_omega_lambda_1}  and \ref{fig:quantum_bt_vs_omega_lambda_05}, where $t_{qb}$ is defined according to the already discussed criterion (\ref{eq:qunatum_breaking_critetion}). Independently of the coupling constant, two different features can be observed as the charge is varied. In the limit of small charge, displayed on the left side of Fig. \ref{fig:quantum_bt_vs_omega_lambda_1}  and \ref{fig:quantum_bt_vs_omega_lambda_05}, the quantum break time grows asymptotically.  This is because, as $\omega$ decreases, we approach the infinitely long lived uncharged vacuum. In this limit, as the collective coupling becomes miniscule, the collective oscillatory period is closer and closer to that of free zero mode particles. Correspondingly, the propagator on top of the condensate is approximately the same as in the free theory one and the system basically consists of diluted quasi-free bosons on top of the unbroken vacuum.
Equivalently, one can understand this feature looking at the diagrammatic behaviour. Out of the two diagrams at $\mathcal{O}(\hbar^2)$, one is simply proportional to the coupling constant (we restore $m$ in this part for clarity)
\begin{equation}\label{eq:local_2l_diagr}
\centering
\parbox{30mm}{
\begin{fmffile}{localself}
\begin{fmfgraph}(70,70)
\fmftop{t0,t1}
\fmfbottom{b0,b1}
\fmf{phantom}{t0,v1,t1}
\fmf{phantom}{b0,v1,b1}
\fmffreeze
\fmf{plain,tension=0.8,right=270}{v1,v1}
\fmf{plain,tension=0.8,right}{v1,v1}
\fmfdot{v1}
\end{fmfgraph}
\end{fmffile}}\sim~~ \frac{\lambda}{m^2}.
\end{equation}
Because the above is independent of the field expectation value, it is not the one responsible for the interaction between the degrees of freedom of the effective action, and, therefore, it is irrelevant for the breaking of the system. The opposite is true for the second diagram, which is proportional to $\lambda^2\phi_{a}^{2}$, non local, and accounts for the interaction between $\phi$ and $G$. Therefore, its relevance for the dynamics heavily depends on the value of $\omega$. In particular, in the limit of small charge (or $\omega \approx m$), its cumulative effects leading to quantum breaking are additionally suppressed

\begin{equation*}
\parbox{20mm}{
\begin{fmffile}{omega_smaller_than_m}
\begin{fmfgraph}(60,60)
\fmfleft{i}
\fmfright{o}
\fmf{phantom,tension=5}{i,v1}
\fmf{phantom,tension=5}{v2,o}
\fmf{plain,left,tension=0.4}{v1,v2,v1}
\fmf{plain}{v1,v2}
\fmfdot{v1,v2}
\end{fmfgraph}
\end{fmffile}
}
\sim \frac{\lambda^2}{m^4} \phi_{a}^{2} = 4 \frac{\lambda}{m^2}\frac{\omega^2-m^2}{m^2} \xrightarrow{\omega-m \ll m} 4 \frac{\lambda}{m^2}\frac{2(\omega-m)}{m} \ll \frac{\lambda}{m^2}.
\end{equation*}
In fact, in the exact free theory limit, this diagram vanishes and the free theory propagators are recovered (with a self-energy like correction coming from the first diagram), and, due to the absence of a non local diagram, no interesting dynamics leading to quantum breaking is observed at this loop order. Hence this explains why $t_{qb}$ increases as $\omega$ decreases. Interestingly, but not surprisingly, $t_{qb}$ grows also in the opposite, big charge, limit ($\omega \gg m$) as it can be seen on the right side of Fig. \ref{fig:quantum_bt_vs_omega_lambda_1}  and \ref{fig:quantum_bt_vs_omega_lambda_05}. In this case, diagrammatically, the non local diagram seems to be enhanced; we have in fact:
\begin{equation*}
\parbox{20mm}{
\begin{fmffile}{omega_bigger_than_m}
\begin{fmfgraph}(60,60)
\fmfleft{i}
\fmfright{o}
\fmf{phantom,tension=5}{i,v1}
\fmf{phantom,tension=5}{v2,o}
\fmf{plain,left,tension=0.4}{v1,v2,v1}
\fmf{plain}{v1,v2}
\fmfdot{v1,v2}
\end{fmfgraph}
\end{fmffile}}
\sim \frac{\lambda^2}{m^4} \phi_{a}^{2} = 4 \frac{\lambda}{m^2}\frac{\omega^2-m^2}{m^2} \xrightarrow{\omega\rightarrow+\infty} 4 \frac{\lambda}{m^2}\frac{\omega^2}{m^2} \gg \frac{\lambda}{m^2},
\end{equation*}

Correspondingly, one would expect quantum breaking to happen even faster. This puzzle can be explained in the following way. The internal degrees of freedom in this case are very different from the free ones, due to the large charge and energy density. Consequently, one should also take into account the scaling of the propagator, and not just the field insertion as done in the above estimate. This is hard to do analytically, however we see this contribution to be small numerically. Moreover, the behaviour in the big charge regime is fully justified if one takes into account the fact that, in this limit, the system becomes highly classical, therefore ensuring a longer breaking. 

Because $t_{qb}$ grows in the opposite limits of small and large charge, there is a value of $Q$ for which breaking happens the fastest. 

\begin{figure}
    \centering
    \begin{subfigure}[b]{0.47\textwidth}
        \includegraphics[width=\textwidth]{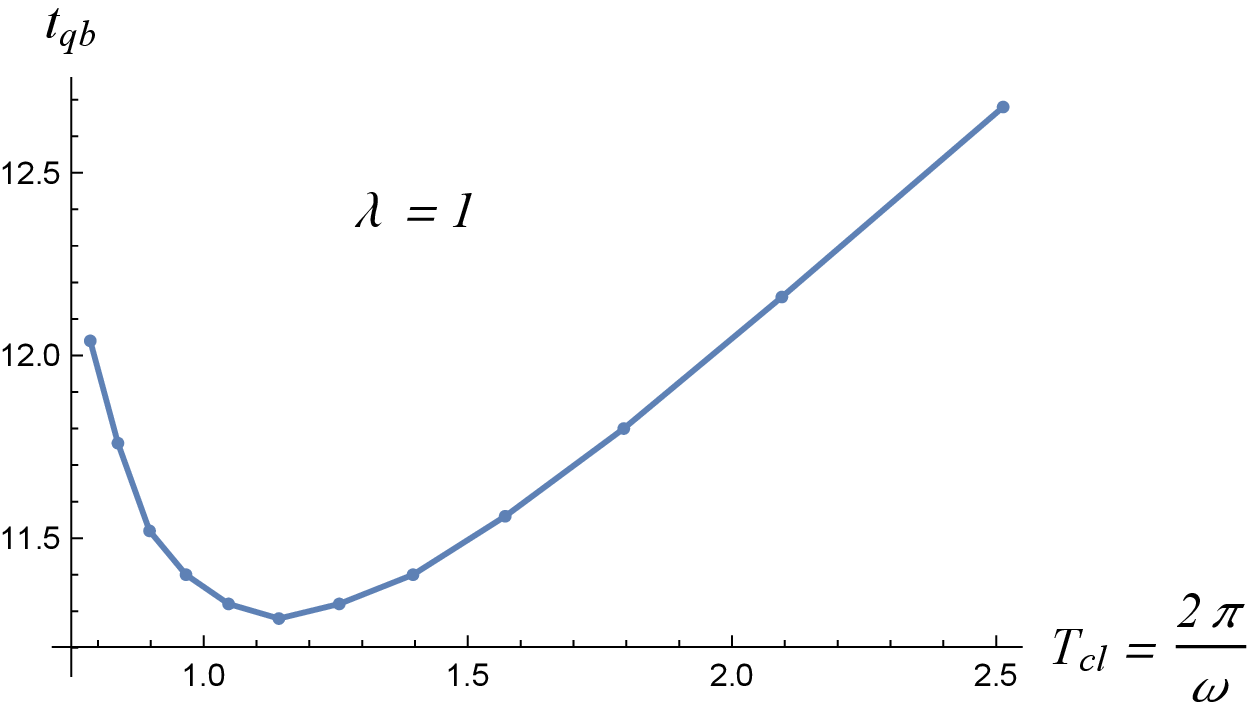}
        \caption{$\lambda = 1$}
        \label{fig:quantum_bt_vs_Tcl_lambda_1}
    \end{subfigure}
    ~
    \centering
    \begin{subfigure}[b]{0.47\textwidth}
        \includegraphics[width=\textwidth]{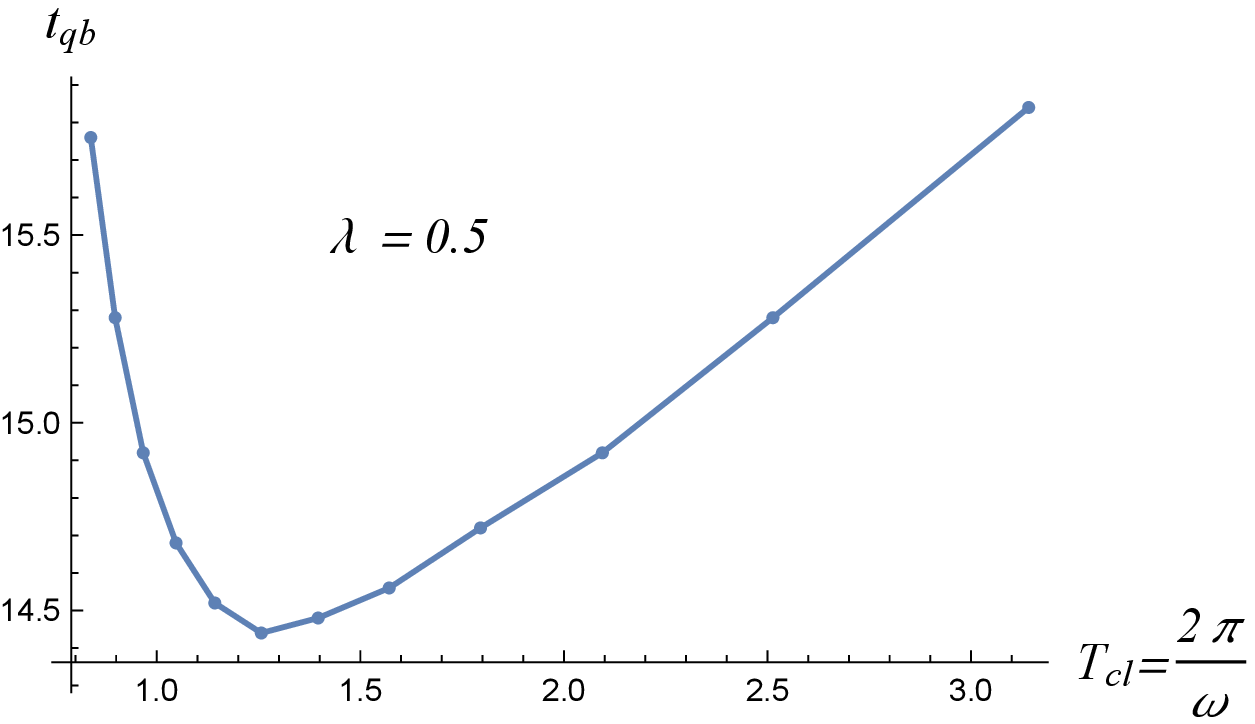}
        \caption{$\lambda = 0.5$}
        \label{fig:quantum_bt_vs_Tcl_lambda_05}
    \end{subfigure}
    \caption{Quantum break-time as a function of period of initial classical oscillations}
    \label{fig:quantum_break-time_vs_Tcl}
\end{figure}

Let us investigate this point in more details. In Fig. \ref{fig:quantum_bt_vs_omega_lambda_1} and \ref{fig:quantum_bt_vs_omega_lambda_05}, the quantum break time is shown for the same set of frequencies, but for different charges (as the couplings are different). The minimum lies at two different charge values: at $Q \sim 6 \cdot 10^3$ and at $Q\sim 10^4$ respectively. However, if we check the dependence of the minimum on a more natural quantity, which is the classical period time $T_{cl}=2\pi/\omega$ as in Fig. \ref{fig:quantum_bt_vs_Tcl_lambda_1} and \ref{fig:quantum_bt_vs_Tcl_lambda_05}, it is evident that the minimum occurs at approximatively the same value i.e. $T_{cl}\sim m^{-1}$.

The attentive reader might wonder whether or not this minimum, as well as the big charge behaviour is indeed due to the high classicality of the system or to a failure of the perturbative $\hbar$ expansion. The latter turns out not to be the case for the following reason. For the given $\omega$ range, the qualitative behaviour we observe is the same as long as the coupling strength is kept weak, i.e.  $\lambda \leq 1$. However, as $\lambda \geq 1$, the dependence was checked to be totally different and, in particular, criterion dependent. If we define the quantum breaking criterion by the ratio $a=Q_q(t_{qb})/Q_{cl}(t_{qb})$ in the region where $Q_{cl}(t)$ monotonically decreases (and $a \in (Q_q(0)/Q_{cl}(0),1]$\footnote{when $a=1$ we recover (\ref{eq:})}), the qualitative behaviour of $t_{qb}$ as a function of $\omega$ changes dramatically with $a$ in the strong coupling regime. Contrarily, in the small coupling regime, the dependence is qualitatively unaffected by the choice of $a$. 

\begin{figure}
    \centering
        \includegraphics[width=\textwidth]{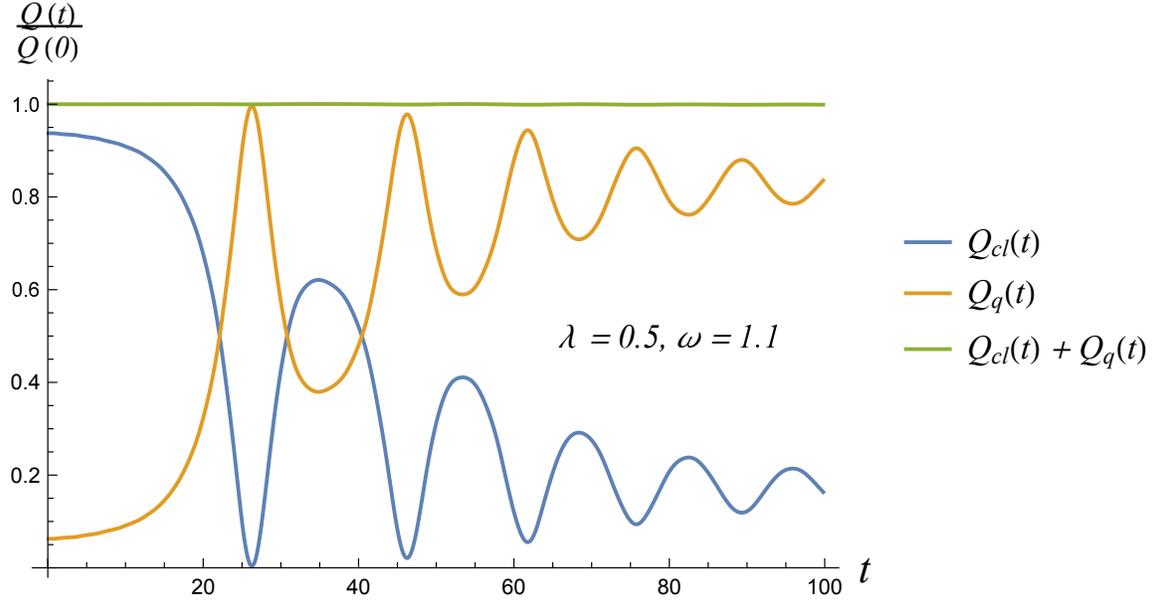}
        \caption{Behaviour of the classical and quantum charge for long time. Conservation of charge is showed. Here $\omega = 1.1 $, $\lambda = 0.5$ .}
        \label{fig:long_time}
\end{figure}

Finally, the long-time dynamics is displayed in Fig. \ref{fig:long_time}. As one can see, the evolution eventually becomes stationary, redistributing the total charge between $\phi$ and $G$. It follows that, asymptotically, the solution approaches values far from the  initial 1-loop stationary solution. Indeed we would like to stress that the evolution is reliable up to $\mathcal{O}(\hbar^3)$ corrections. Since such effects, in principle, could be cumulative, it is unclear whether or not the plotted solution is reliable up to the very end of the simulation.

\subsection{Comparison with the perturbative expansion}
\begin{figure}
    \centering
        \includegraphics[width=\textwidth]{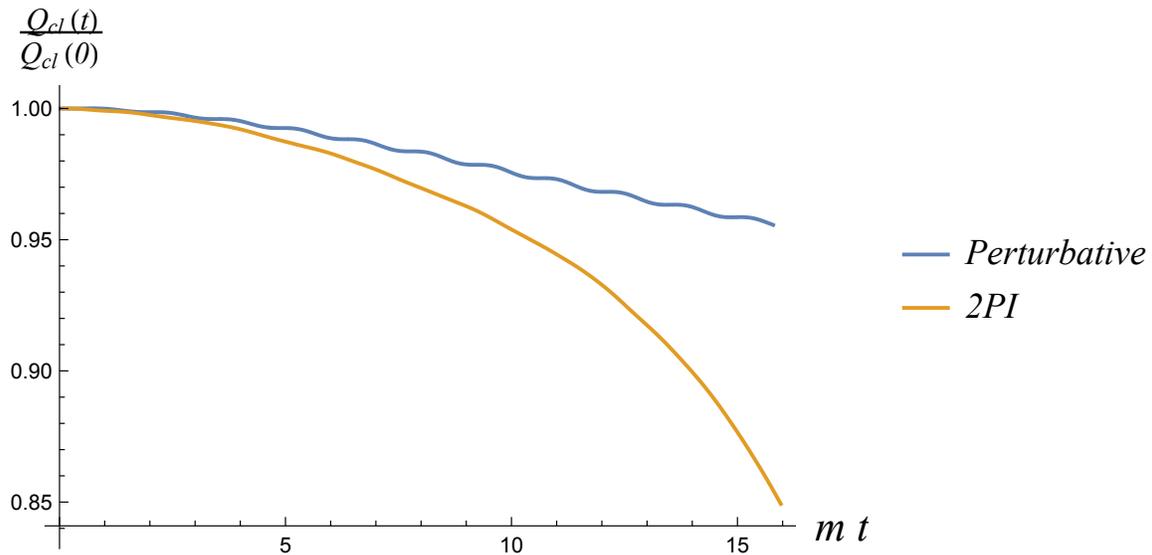}
        \caption{Perturbative expansion compared with fully resumed 2PI for $\lambda=m^2$ and $\omega= 1.1\, m$.}
        \label{fig:pert_vs_loop}
\end{figure}

The perturbative equations of motion in the coupling $\lambda$ can be easily derived using the Schwinger-Keldysh formalism \cite{Schwinger:1960qe,Keldysh:1964ud} as done in \cite{Boyanovsky:1994me} for a real scalar field. To do so, we define a density matrix $\rho$ describing the system at $t=0$ evolving according to (in the interaction picture)
\[
 i \frac{\partial \rho (t)}{\partial t} = \left[H_I(t),\rho(t) \right],
\]
whose formal solution is 
\[
\rho (t) = U_I(t,t_0)\rho(t_0) U_I^{-1}(t,t_0) ,
\]
with $U_I(t,t_0) = {T_\mathcal{C}} \exp{\left(-i\int_0^t d t' H_I(t')\right)} $; $T_\mathcal{C}$ denoting the ordering with respect to the in-in contour.\\
The expectation value of the field is then given by
\begin{equation}
    \phi(t) = \langle \Phi (t)\rangle = \frac{\text{Tr}\, \Phi \rho(t)}{\text{Tr}\,\rho(t)} = \text{Tr}\left[ \rho(t_0)\, T_{\mathcal{C}}\left( \Phi(t)^{+} \exp\left[ -i \int_0^t \left[H_I(t')^{+} - H_I(t')^{-}\right]\right] \right)  \right] ,
\end{equation}
where $\pm$ indicates on which branch the operators are to be evaluated.\\
The generating functional thus becomes
\begin{equation}
    Z[J^+,J^-]= \int d \phi^{+} d \phi^{-} \rho(\phi^{+},\phi^{-};t_0)\int_{bc} \mathcal{D} \Phi^{+} \mathcal{D}\Phi^- \exp\left[ {i}\, S[\Phi^{+},\Phi^{-},J^{+},J^{-}] \right],
\end{equation}
where the boundary conditions correspond to $\Phi^+(0,x)= \phi^+(x)$, $\Phi^-(0,x)=\phi^-(x)$ and $\Phi^+(t,x)=\Phi^-(t,x)$ and 
\[
S[\Phi^{\pm},J^{\pm}]= \int_0^t dt' \int dx \left[\mathcal{L}[\Phi^{+}] - \mathcal{L}[\Phi^{+}] + J^{+} \Phi^{+} - J^{-} \Phi^{-} \right]\,.
\]
To find the equations of motion we employ the tadpole method as in \cite{Boyanovsky:1994me,Weinberg:1974hy}. Splitting the field into $\Phi(t,x)^{\pm} = \phi(t) + \delta \phi(t,x)^{\pm}$ where $\delta \phi(t,x)$ is a fluctuation, one can ask for which background $\phi(t)$ the tadpole condition,
\begin{equation}
\label{tadcond}
    \langle \delta \phi(t,x)^{\pm}\rangle =0,
\end{equation}
is satisfied.
Assuming that at $t=0$ the vacuum is the same as in the free theory (i.e. $|0(t=0)\rangle = |0\rangle$) one obtains, from the path integral
\begin{equation}
\label{treelev}
\langle \delta\phi^+\rangle = i \int_0^t dt' \int dx \left[ \left.\frac{\delta S^+}{\delta \Phi^+}\right|_{\phi(t')} G^{++}(t,t';x,y) - \left. \frac{\delta S^-}{\delta \Phi^-}\right|_{\phi(t')} G^{+-}(t,t';x,y) \right] + O(\mathcal{\hbar}),
\end{equation}
where $G^{\pm\pm}(t,x;t',y)= \langle T_{\mathcal{C}}\left(\Phi^\pm(t,x)\Phi^\pm(t',y)\right)\rangle$. In particular $G^{++}, G^{+-}, G^{-+}$ and $G^{--}$ correspond to the Feynmann, the advanced, the retarded and Dyson propagator respectively. We see that in order for \eqref{treelev} to satisfy the tadpole condition \eqref{tadcond} the background field needs to be a stationary point of the classical action. \\
We now evaluate perturbatively in $\lambda$ the tadpole condition for the fluctuations of \eqref{eq:action}, whose action is given by (from now on we restore $m$)
\begin{equation}
\begin{split}
    S[\delta \phi^{+} , \delta \phi^{-}] = \int dt\, dx &\,\left[\frac{\left(\partial_\mu \phi_a^{+} \right)^2}{2} - \frac{m^2 (\delta \phi^+_a)^2}{2} + \delta \phi_a^+\left.\frac{\delta S}{\delta \varphi_a^+}\right|_{\varphi=\phi} - \frac{(\delta \phi_a^+)^2}{2}\left(m^2 + \frac{\lambda}{4} \phi_a^2\right) -\right.\\ 
    &\left. \frac{\lambda}{4} \delta \phi_a^+ \delta \phi_b^+\phi_a \phi_b - \frac{\lambda}{16}\left[(\delta \phi_a^+)^4 + \phi_a^4\right] - \frac{\lambda}{4} \delta\phi_a^+ \delta \phi_b^2 \phi_a - \left(\delta \phi^+ \leftrightarrow \delta \phi^-\right)\right].
    \end{split}
\end{equation}
Up to $\lambda^2$, the tadpole condition schematically reads:

\begin{equation}
\parbox{20mm}{
\begin{fmffile}{pert1}
\begin{fmfgraph}(40,40)
\fmfleft{i}
\fmfright{o}
\fmf{plain,label=lol}{i,v}
\fmf{phantom}{v,o} 
\fmfv{decor.shape=square,decor.filled=full,decor.size=2thick}{v}
\end{fmfgraph}
\end{fmffile}}
 + \quad
\parbox{20mm}{
\begin{fmffile}{pert2}
\begin{fmfgraph}(40,40)
\fmfleft{i}
\fmfright{o}
\fmf{plain}{i,v}
\fmf{dashes}{v,o}
\fmf{plain}{v,v}
\fmfdot{v}
\end{fmfgraph}
\end{fmffile}
}
 + \quad
\parbox{20mm}{
\begin{fmffile}{pert3}
\begin{fmfgraph}(40,40)
\fmfleft{l1,l2}
\fmfright{r1,r2}
\fmf{plain}{l1,v1}
\fmf{dashes}{l2,v1}
\fmf{dashes}{r1,v2}
\fmf{dashes}{r2,v2}
\fmf{plain,left=1.,tension=0.5}{v1,v2,v1}
\end{fmfgraph}
\end{fmffile}
}
=0,
\end{equation}

where the dashed line corresponds to an insertion of the background field $\phi_a$. The first diagram is nothing but the tree level equation of motion as one can see from \eqref{treelev}. The second diagram can be absorbed by a mass renormalization. Altogether they lead to:
\begin{equation}
\label{perteom}
\left\lbrace\begin{array}{l}
   \displaystyle \Ddot{\phi}_1 + m^2 \phi_1 + \lambda (\phi_1^2 + \phi_2^2)\phi_1 - \frac{\lambda^2}{2} \phi_1 \int_0^t dt' \left(\phi_1(t')+ \phi_2(t')\right)^2 \int \frac{dk}{2\pi} \frac{\sin\left[ 2 \omega_k(t-t')\right]}{2 \omega_k^2} =0 \\
   \\
   \displaystyle \Ddot{\phi}_2 + m^2 \phi_2 + \lambda (\phi_1^2 + \phi_2^2)\phi_2 - \frac{\lambda^2}{2} \phi_2 \int_0^t dt' \left(\phi_1(t')+ \phi_2(t')\right)^2 \int \frac{dk}{2\pi} \frac{\sin\left[ 2 \omega_k(t-t')\right]}{2 \omega_k^2} =0 
    
\end{array}\right..
\end{equation}
where the renormalization of the mass has already been taken into account by dropping the second diagram. Note that the choice of the time contour ensured the last term to be causal. 
\\
The comparison between the fully resummed 2-loop dynamics and the perturbative analysis to $\lambda ^2$ is shown in Fig. \ref{fig:pert_vs_loop}. To fix the initial conditions for the perturbative equation of motion \eqref{perteom} we considered the saddle point solution.

As it can be seen, the leading behaviour for small times is similar. However, at later times, the 2PI solution breaks faster. For longer times (although not shown explicitly), the perturbative solution oscillates around $Q_{cl}(0)$. Clearly, on such time scales, the perturbative solution is no longer reliable and  the leading behaviour of the breaking is captured by the 2PI effective action. \\

\section{Conclusions and outlook}
In this work we explored how quantum effects dynamically change the evolution of a relativistic BEC composed from a charged scalar field with repulsive quartic self interaction. In particular, we saw how such a system experiences quantum breaking deviating from the solution given by stationary point of the 1PI effective action. To do so, we have computed the 2PI effective action up to two loops. The main advantage of this approach is that both the 1 and 2-point connected correlation functions are treated as interacting degrees of freedom. Thus, it allowed us to formulate a natural criterion for quantum breaking (\ref{eq:qunatum_breaking_critetion}) in terms of classical $Q_{cl}(t)$ and quantum $Q_{q}(t)$ charges defined in (\ref{eq:classical_charge}) and (\ref{eq:quantum charge}) by comparing how the total charge $Q$, which is the conserving sum of the two (\ref{eq:general_charge}), is redistributed among them. Therefore, given an initial configuration of highly occupied zero modes modelling the stationary solution to 1PI effective action, we observed how rescattering effects slowly break the configuration, which is fully stable from a classical point of view. This breaking is shown in Fig. \ref{fig:Qcl(T)for_diff_omega} and \ref{fig:Qq(T)for_diff_omega} exposing mutual change of the two charges, while total charge is conserved as it depicted in Fig. \ref{fig:Qcl_and_full_charges}. Although it was not possible to extract explicitly the functional dependence of the breaking, interesting features were observed. It turned out that in two opposite limits $Q\rightarrow 0$ and $Q\rightarrow \infty$ the quantum break time grows as one can see in Fig. \ref{fig:quantum_bt_vs_omega_lambda_1} and \ref{fig:quantum_bt_vs_omega_lambda_05}. As explained earlier, this is due to two different trends. For small charge the configuration is closer to its true vacuum and, correspondingly, the effects breaking the configuration are suppressed. In the big charge limit, the theory is highly classical and, consequently, more rescattering events  are necessary for the effects to be dynamically relevant. The minimal quantum break-time between these two regimes was found. The point where quantum break-time is minimal seems to be almost independent of the coupling constant and happens at $T_{cl}\simeq m^{-1}$ or $\omega \simeq 2\pi m$ as depicted in Fig. \ref{fig:quantum_bt_vs_Tcl_lambda_1} and \ref{fig:quantum_bt_vs_Tcl_lambda_05}. 

To show the necessity of using the 2PI effective action resummation, we compared our findings with the perturbative expansion in the coupling constant. Both approaches show a similar behaviour for small times, but dramatically differ later (Fig. \ref{fig:pert_vs_loop}). In particular, only the $\hbar$ expansion happens to capture the breaking, as the perturbative one leads only to an oscillatory behaviour of the classical charge. Clearly, this is because the first resums a finite amount of different infinite subsets of diagrams while the latter accounts only for the first two contributions. 

In view of our results, some interesting remarks can be made. First of all, we question the saddle point approximation used in most of the computations of a BEC condensate when considering  processes which have a typical timescale (or lengthscale) of order of the breaking we observe. This might lead to erroneous conclusions as quantum breaking is jeopardising the reliability of the background. Moreover, we believe that the corrections we observe might be relevant, and should be taken into account in order to better approximate the quantum evolution of physical condensates. It is worth mentioning that within a full proper treatment, quantum and thermal fluctuations appear on equal footing. Hence quantum effects are dominant only if thermal fluctuations are negligible. If this is not the case, a statistical description of the system must be taken into account.

 Secondly, we note that the breaking timescale is generically longer than \eqref{eq:quantumbreaktime}. This is not in a contradiction with \cite{Dvali:2017eba}, as they give a lower bound for this quantity while, in our approach, the effects of  backreaction are observed explicitly in real time. Since in this framework quantum breaking really looks like a reorganisation of the internal degrees of freedom of the system, it is natural to explore its relation with the so-called scrambling effect. In particular, for an attractive interaction, the appearance of a classical instability corresponds to the presence of a solitonic configuration in the spectrum 
 This has already been explored in the non relativistic case in ref. \cite{Dvali:2013vxa}. It was found that the scrambling time scales as
\begin{equation}
\label{eq:scr}
    t_s \sim \gamma^{-1}\log(N),
\end{equation}
where $N$ is the number of constituents and $\gamma$ the Lyapunov exponent. Because the scrambling time is classically infinite (as $N\propto \hbar^{-1}$), it follows that \eqref{eq:scr}  has a deep quantum origin. Due to the presence of classical instabilities, the breaking of the configuration is exponentially fast, and, as a consequence of quantum effects, is additionally amplified by $\log N$. Moreover, the authors of ref. \cite{Dvali:2013vxa} claim that \eqref{eq:scr} is a consequence of the connection between quantum breaking and the scrambling of information. Therefore, verifying this effect within the 2PI framework would not only give foundation to this last statement but also verify such behaviour in the relativistic regime.

\paragraph{Acknowledgement} We are thankful to Gia Dvali, Georgios Karananas and Emin Nugaev for  fruitful discussions, reading the manuscript and useful comments.
\newpage
\appendix
\section{1PI effective action of BEC at one loop}\label{sec:1PI_solution}
In this section we will derive initial conditions for the statistical part of the Green's function and field.
As already explained, we set as initial conditions the functions extracted from the 1-loop approximation of the effective action. Namely, we are going to solve here equations (\ref{eq:1loop_stationary_conditions}). Within this order the evolution of the system is trivial because the first loop contains only a local self-energy contribution. It follows that the solution is stationary.

To find the condensate solution we use an ansatz similar to (\ref{eq:classical_ansatz}) together with an ansatz for the Green's function as well

\begin{equation}\label{eq:}
\left\lbrace\begin{array}{l}
     \displaystyle \phi_a (t,x) = R_{ab}(\omega t) f_{b}  \\
     \\
    \displaystyle G_{ab}(t,x;\tau, y) =   R_{ac}(\omega t)R_{bd}(\omega \tau) \Tilde{G}_{cd}(t-\tau,x-y) 
\end{array}\right.
\end{equation}
where $f_a$ is a constant vector and $R_{ab}(\theta) \in SO(2)$ was defined in (\ref{eq:2d_rotation_matrix}).

One can actually see that after this change of variables both the 1 and 2PI effective action admit a homogeneous nontrivial solution for the field expectation value while still preserving time and space translations as explicit time dependence fully factors out.

Plugging the new variables in the system (\ref{eq:1loop_stationary_conditions}) we get

\begin{equation}\label{eq:1loop_st_cond_2}
 \left\lbrace   \begin{array}{l}
          \displaystyle \left(\left(\omega^2+\gamma^2 - p_{n}^{2} - m^2 -\frac{\lambda}{4}f_{d}^{2} \right)\delta_{ac} - 2 i \omega\gamma \epsilon_{ac} - \frac{\lambda}{2}f_{a}f_{c}\right)\Tilde{G}_{cb}(\gamma,p_{n}) = i\delta_{ab}\\
          \\
         \displaystyle \left(\left(-\omega^2 + m^2 +\delta m^2 +\frac{\lambda}{4}f_{d}^{2}\right)\delta_{ab} +\frac{\lambda}{4}\left(\vphantom{\left(G\right)^2}\Tilde{G}_{dd}(0,0)\delta_{ab} + 2\, \Tilde{G}_{ab}(0,0)\right)  \right)f_{b}=0 
    \end{array}\right.
\end{equation}
where 
\begin{equation}\nonumber
    \Tilde{G}_{ab}(t-\tau,x-y) = \frac{1}{L}\int \frac{d\gamma}{2\pi}\sum_{n=-\infty}^{+\infty}e^{-i\gamma (t-\tau)+i p_n (x-y)}\left.\Tilde{G}_{ab}(\gamma,p_n)\right|_{ p_n=\frac{2\pi n}{L}}
\end{equation}
and
\begin{equation}
    \delta m^2 = -\lambda\sum_{n=-\infty}^{+\infty}\frac{1}{2\sqrt{p_{n}^{2}+m^2}}
\end{equation}
is the counterterm taken from the free theory, which is natural since the true vacuum of the theory is intact and well defined.

The solution to these equations is actually a stationary condition for the Coleman-Weinberg potential of the Bose-Einstein condensate. The first equation defines a Green's function at the top of the condensate and the second one is giving an expectation value. The  $\mathcal{O}(\hbar^0)$ part of the second equation are just the classical equations of motions while the $\mathcal{O}(\hbar)$ gives the 1-loop correction.

To solve the system (\ref{eq:1loop_stationary_conditions}) we have to understand when the linear operator acting on $\Tilde{G}$ in the 1-st eq. of (\ref{eq:1loop_st_cond_2}) is invertible. It turns out that this is the case when the following condition is fulfilled
\begin{equation}\label{eq:invertibility_condition}
    \sqrt{f_{a}^{2}} > v 
\end{equation}
where  $v$ is defined in (\ref{eq:modulus}). Thus, we see that actually the modulus of the field must be bigger than its saddle-point value. Next, keeping in mind this condition, we can invert $\Tilde{G}_{0}^{-1}$ and derive
\begin{equation}\label{eq:G_1loop_in_momentum_space}
    \Tilde{G}_{ab}(\gamma,p_n) = \frac{i\left(\displaystyle\left(\displaystyle \omega^2+\gamma^2 - p_{n}^{2} - m^2 -\frac{\lambda}{4}f_{d}^{2} \right)\delta_{ac} + 2 i \omega\gamma \epsilon_{ac} - \frac{\lambda}{2}f_{a}f_{c}\right)}{\displaystyle\left(\vphantom{\frac{\lambda}{4}f_{d}^{2}}\gamma^2-\gamma_{+}^{2}(p_{n}) + i 0\right)\left(\gamma^2 - \gamma_{-}^{2}(p_{n})+ i 0 \vphantom{\frac{\lambda}{4}f_{d}^{2}}\right)}
\end{equation}
where $i0$ indicates the usual Feynman boundary conditions, $\epsilon_{ab}$ is 2-dimensional Levi-Civita tensor and the excitation energies are defined by
\begin{equation}
    \gamma_{\pm}^{2} = p_{n}^2 +\omega^2+m^2 +\frac{\lambda}{2} f^2 \pm \sqrt{\displaystyle 4\left(p_{n}^2+m^2\right)\omega^2 + 8 \lambda f^2 \omega^2 +\frac{1}{4}\lambda^2 f^4}
\end{equation}

\begin{figure}
  \centering
    \includegraphics[width=0.5\textwidth]{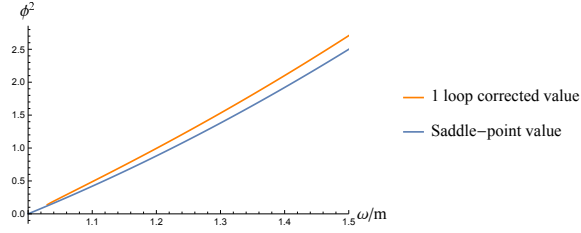}
    \caption{Field values for saddle-point and 1-loop approximations for $\lambda = 1$, $m=1$, $L=10\, m^{-1}$}
    \label{fig:initial_modulus}
\end{figure}

Plugging (\ref{eq:G_1loop_in_momentum_space}) in the second eq. of (\ref{eq:1loop_st_cond_2}) we eventually find the field expectation value $f_a$. Apparently, this equation depends only on the modulus of the field, therefore, for convenience we choose 
\begin{equation}
    f_a = \left(\begin{array}{l}
         f  \\
         0 
    \end{array}\right).
\end{equation}
Thus the stationary condition for the field becomes
\begin{equation}\label{eq:1loop_modulus_equation}
     -\omega^2 + m^2  +\frac{\lambda}{4}f^{2} +\left(\frac{3 \lambda}{4}\Tilde{G}_{11}(0,0) + \frac{ \lambda}{4}\Tilde{G}_{22}(0,0) + \frac{\lambda}{2} \Tilde{G}_{12}(0,0) +\delta m^2 \right) = 0.
\end{equation}
By solving this numerically we can find the modulus of the field and compare this with the saddle-point value. This comparison is presented in Fig. \ref{fig:initial_modulus} for specific parameters. We took here $\lambda = m^2=1$ for illustrative purposes, though, one can see that even in this case the corrections to the modulus are quite small. One can also observe that in the 1-loop case, for some $\omega > m$, a solution to equation (\ref{eq:1loop_modulus_equation}) does not exist. This is in contrast with the classical case which admits a solution for all frequencies bigger than the mass of the free boson $m$.

Another point which is worth mentioning is the invertibility of the operator in (\ref{eq:1loop_st_cond_2}) acting on the Green's function. We have made an a priory assumption about it's invertibility given by (\ref{eq:invertibility_condition}) and now explicitly checked that for the solutions of (\ref{eq:1loop_modulus_equation}) this inequality is indeed fulfilled.

The last step is to extract the initial conditions for $F_{ab}(0,x;0,y)$ from $G^{0}_{ab}(x,y)$. We easily achieve this by integrating over $\gamma$ in (\ref{eq:fourier_of_1loop_GF}). After that, we use the decomposition formula (\ref{eq:GF_decomposition}) (but with Feynman contour not Schwinger-Keldysh) and extract $F^{0}_{ab}(x,y)$
\begin{eqnarray}
    F^{0}_{ab}(t,x;\tau,y) = \frac{1}{L}\sum_{n=-\infty}^{+\infty} R(\omega t)\left( \frac{1}{4\gamma_{+}}\left( e^{-i\gamma_{+} (t-\tau)}\Tilde{G}(\gamma_{-},p_{n}) + e^{i\gamma_{+} (t-\tau)}\Tilde{G}(-\gamma_{+},p_{n}) \right) - \right. \nonumber
    \\
    \left. \frac{1}{4\gamma_{-}}\left( e^{-i\gamma_{-} (t-\tau)}\Tilde{G}(\gamma_{-},p_{n}) - e^{i\gamma_{-} (t-\tau)}\Tilde{G}(-\gamma_{-},p_{n}) \right) \right)R^{T}(\omega \tau)\frac{e^{i\,p_{n}(x-y)}}{\left(\gamma_{+}^{2}-\gamma_{-}^{2}\right)}.
\end{eqnarray}
Setting $t=\tau=0$ we get an initial condition for $F_{ab}(x,y)$. Also, we can extract $\rho^{0}_{ab}(x,y)$ and check explicitly that it satisfies the canonical commutation relations (\ref{eq:initial_conditions_for_rho}).
\section{Loop diagrams}\label{sec:non_loc_self_energy}
In this part of the appendix we will explicitly write down the 2-loop contributions to self energy expressed in terms of the Green's function decomposition (\ref{eq:G_decomp}). The first diagram is
\begin{eqnarray}
\parbox{30mm}{
\begin{fmffile}{1stdiagr}\begin{fmfgraph}(70,70)
\fmftop{t0,t1}
\fmfbottom{b0,b1}
\fmf{phantom}{t0,v1,t1}
\fmf{phantom}{b0,v1,b1}
\fmffreeze
\fmf{plain,tension=0.8,right=270}{v1,v1}
\fmf{plain,tension=0.8,right}{v1,v1}
\fmfdot{v1}
\end{fmfgraph}
\end{fmffile}}
= ~~~ -\frac{ \lambda}{16}\int_{\mathcal{C}} dx\left( 2 G_{ab}(x,x) G_{ab}(x,x) + \left(G_{aa}(x,x)\right)^2 \right). 
\end{eqnarray}
As one can see this diagram is integrated over a single space-time point, therefore it contributes only to the local self-energy of the field, thus providing a trivial correction to the evolution.

The second diagram is
\begin{eqnarray}
\parbox{20mm}{
\begin{fmffile}{2nddiagram}
\begin{fmfgraph}(60,60)
\fmfleft{i}
\fmfright{o}
\fmf{phantom,tension=5}{i,v1}
\fmf{phantom,tension=5}{v2,o}
\fmf{plain,left,tension=0.4}{v1,v2,v1}
\fmf{plain}{v1,v2}
\fmfdot{v1,v2}
\end{fmfgraph}
\end{fmffile}}
=~~~ \frac{i \lambda^2}{16}\int_{\mathcal{C}} dx \int_{\mathcal{C}} dy \left(\vphantom{\left(G\right)^2}G_{ab}\left(x,y\right)\phi_{a}(x)\phi_{b}(y) G_{cd}(x,y)G_{cd}(x,y) \,+\right. \\
    \left.2\, \phi_{a}(x) G_{ab}(x,y) G_{bc}(x,y)G_{cd}(x,y) \phi_{d}(y) \vphantom{\left(G\right)^2}\right).
\end{eqnarray}
This one depends explicitly on the field expectation value, probes two different space-time points and correspondingly contributes non trivially to the evolution.

The sum of these diagrams is the $\mathcal{O}\left(\hbar^2\right)$ part of the effective action $\Gamma_2 [ \phi, G ]$.
Plugging this contributions into (\ref{eq:stationary_conditions}) we obtain
\begin{equation}\label{eq:eom_reduced}\left\lbrace\begin{array}{l}
     \displaystyle -\frac{\delta S [\phi]}{\delta \phi_{a}(x)} + \frac{ \lambda}{4}\left(G_{dd}(x,x)\delta_{ab}+2 G_{ab}(x,y)\right)\phi_{b}(x) = \frac{\delta \Gamma_2 [\phi,G]}{\delta\phi_{a}(x)}  \\
     \\
     \displaystyle \left(\partial^{2}\delta_{ac} + M_{ac}^{2}(x)\right)G_{cb}(x,y) +  i \int dz \, \Sigma_{ac}^{non-local}(x,z)G_{cb}(z,y) = - i\delta_{ab}\delta \left(x-y\right) 
\end{array}
    \right.
\end{equation}
where
\begin{equation}
  \frac{\delta \Gamma_2 [\phi,G]}{\delta\phi_{a}(x)} = \frac{i\lambda^2}{8}\int_{\mathcal{C}} dy \left( \vphantom{\left(G\right)^2} G_{cd}(x,y)G_{cd}(x,y) G_{ab}(x,y)\phi_{b}(y) + 2\, G_{ab}(x,y)G_{cb}(x,y)G_{cd}(x,y)\phi_{d}(y) \right)
\end{equation}
and
\begin{equation}
    M_{ab}^{2}(x) = \frac{\lambda}{4} \left(\vphantom{\left(G\right)^2}\phi_{c}^{2}(x) +  G_{cc}(x,x) \right)\delta_{ab} + \frac{\lambda}{2} \left(\vphantom{\left(G\right)^2}\phi_{a}(x)\phi_{b}(x) +  G_{ab}(x,x)\right) + \delta m^2\delta_{ab}\nonumber
\end{equation}
\begin{eqnarray*}
  \Sigma_{ac}^{non-local} (x,z) = 
  - \frac{ \lambda^{2}}{8} \left( \vphantom{\left(G\right)^2}G_{df}(x,z)G_{df}(x,z)\phi_{a}(x)\phi_{c}(z) + 2 \phi_{d}(x) G_{df}(x,z)\phi_{f}(z) G_{ac}(x,z) \,+ \right. \\
    2 \, \phi_{a}(x) G_{fc}(x,z) G_{fd}(x,z)\phi_{d}(z) + 2 \phi_{f}(x) G_{fd}(x,z) G_{ad}(x,z)\phi_{c}(z) + \\
   \left. 2 \, \phi_{f}(x) G_{fc}(x,z) G_{ad}(x,z)\phi_{d}(z)\vphantom{\left(G\right)^2}\right) .
\end{eqnarray*}

To derive causal equations of motion we decompose the non-local part of the self energy in the same fashion as the propagator (\ref{eq:GF_decomposition})
\begin{equation}\label{eq:self_energy_decomposition}
    \Sigma_{ab}^{non-local}(x,y) = \Sigma_{ab}^{F}(x,y)-\frac{i}{2}\mathrm{sgn}_{\mathcal{C}}(x^0-y^0) \Sigma_{ab}^{\rho}(x,y)
\end{equation}
Using this two decompositions (\ref{eq:GF_decomposition}) and (\ref{eq:self_energy_decomposition})  we can rewrite (\ref{eq:eom_reduced}) in the form (\ref{eq:main_EOM}).

To get an explicit expression of $\Sigma^{F}$ and $\Sigma^{\rho}$ one can notice that $\Sigma^{non-local}$ is actually a function of two products, namely, $G_{ab}(x,y)G_{cd}(x,y)$ and $\phi_{a}(x)\phi_{b}(y)$
\begin{equation}
    \Sigma^{non-local} (x,y) =  \Sigma^{non-local} \left(G_{ab}(x,y)G_{cd}(x,y) , \phi_{a}(x)\phi_{b}(y)\right).
\end{equation}
Using this property it follows that both parts of the decomposition (\ref{eq:self_energy_decomposition}) can be expressed as
\begin{eqnarray}
    \Sigma_{ab}^{F}(x,y) = &\displaystyle \Sigma^{non-local} \left(F_{ab}(x,y)F_{cd}(x,y) - \frac{1}{4}\rho_{ab}(x,y)\rho_{cd}(x,y) , \phi_{a}(x)\phi_{b}(y)\right) \\
        \Sigma_{ab}^{\rho}(x,y) = & \displaystyle \Sigma^{non-local} \left(\vphantom{\frac{1}{4}}F_{ab}(x,y)\rho_{cd}(x,y) + \rho_{ab}(x,y)F_{cd}(x,y) , \phi_{a}(x)\phi_{b}(y)\right). 
\end{eqnarray}

Finally $\Sigma^{\phi}$ from the first equation of system (\ref{eq:main_EOM}) is
\begin{eqnarray}
    \Sigma^{\phi}_{ab}(x,y) = \frac{\lambda^2}{8}\left(\left(F_{cd}^2(x,y)-\frac{1}{4}\rho_{cd}^2(x,y)\right)\rho_{ab}(x,y) + 2 F_{cd}(x,y)\rho_{cd}(x,y)F_{ab}(x,y) \, + \right.\nonumber\\
    2\left(\rho_{ad}(x,y)F_{cd}(x,y)F_{cb}(x,y) + F_{ad}(x,y)\rho_{cd}(x,y)F_{cb}(x,y) + F_{ad}(x,y)F_{cd}(x,y)\rho_{cb}(x,y)\right) - \nonumber\\
    \left.\frac{1}{4}\rho_{ad}(x,y)\rho_{cd}(x,y)\rho_{cb}(x,y)\right)
\end{eqnarray}
in terms of decomposition (\ref{eq:G_decomp}).

\end{document}